\documentclass[twocolumn]{article}
\usepackage{geometry}
\usepackage{microtype}
\usepackage{graphicx}
\usepackage{booktabs} % for professional tables

\usepackage{hyperref}
\usepackage[ruled]{algorithm2e}

\usepackage{amsmath}
\usepackage{amssymb}
\usepackage{mathtools}
\usepackage{amsthm}

\usepackage[capitalize,noabbrev]{cleveref}

\theoremstyle{plain}

\theoremstyle{definition}

\theoremstyle{remark}

\usepackage[disable,textsize=tiny]{todonotes}
\usepackage{url}
\usepackage{graphicx}
\usepackage{xcolor}
\usepackage{colortbl}
\usepackage{import}
\usepackage{multirow}
\usepackage{wrapfig}
\usepackage{pdflscape}
\usepackage{float}
\usepackage{subcaption}
\usepackage{authblk}
\usepackage{natbib}
\usepackage{libertine}
\usepackage{fancyhdr}
\usepackage{pifont}
\title{Implicit Incompressible Porous Flow using SPH}

\author[1]{Timna Böttcher}
\author[1]{Stefan Rhys Jeske}
\author[1]{Lukas Westhofen}
\author[1]{Jan Bender}

\affil[1]{Computer Animation, RWTH Aachen University, Aachen, Germany}

\date{}

\usepackage{xcolor}

\usepackage{amsmath}
\usepackage{bbold}
\usepackage{subcaption}
\usepackage{multirow}
\usepackage{siunitx}
\usepackage{enumitem}

\newcommand{\eq}[1]{Eq.~\eqref{#1}}
\newcommand{\fig}[1]{Fig.~\ref{#1}}

\renewcommand{\v}[1]{\boldsymbol{\mathbf{#1}}}

\newcommand{\idMat}{\mathbb{1}}

\newcommand{\dt}{\Delta t}

\newcommand{\fpID}{{i}}
\newcommand{\fnID}{{j}}
\newcommand{\spID}{{s}}
\newcommand{\snID}{{t}}

\newcommand{\fpp}[1]{{{#1}_{\fpID}}}
\newcommand{\fpexp}[2]{{{#1}_{\fpID}^{#2}}}
\newcommand{\fn}[1]{{{#1}_{\fnID}}}
\newcommand{\fnexp}[2]{{{#1}_{\fnID}^{#2}}}
\newcommand{\influidneighborhood}[1]{\fnID \in \mathcal{N}_{{#1}}^F}

\newcommand{\spp}[1]{{{#1}_{\spID}}}
\newcommand{\spexp}[2]{{{#1}_{\spID}^{#2}}}
\newcommand{\sn}[1]{{{#1}_{\snID}}}
\newcommand{\snexp}[2]{{{#1}_{\snID}^{#2}}}
\newcommand{\insolidneighborhood}[1]{\snID \in \mathcal{N}_{{#1}}^S}

\newcommand{\Wff}{{W_{\fpID \fnID}}}
\newcommand{\Wfs}{{W_{\fpID \snID}}}
\newcommand{\Wss}{{W_{\spID \snID}}}
\newcommand{\Wsf}{{W_{\spID \fnID}}}

\begin{document}

\twocolumn[
    \begin{@twocolumnfalse}
        \maketitle

        \begin{abstract}
We present a novel implicit porous flow solver using SPH, which maintains fluid incompressibility and is able to model a wide range of scenarios, driven by strongly coupled solid-fluid interaction forces.
Many previous SPH porous flow methods reduce particle volumes as they transition across the solid-fluid interface, resulting in significant stability issues.
We instead allow fluid and solid to overlap by deriving a new density estimation.
This further allows us to extend SPH pressure solvers to take local porosity into account and results in strict enforcement of incompressibility.
As a result, we can simulate porous flow using physically consistent pressure forces between fluid and solid.
In contrast to previous SPH porous flow methods, which use explicit forces for internal fluid flow, we employ implicit non-pressure forces.
These we solve as a linear system and strongly couple with fluid viscosity and solid elasticity.
We capture the most common effects observed in porous flow, namely drag, buoyancy and capillary action due to adhesion.
To achieve elastic behavior change based on local fluid saturation, such as bloating or softening, we propose an extension to the elasticity model.
We demonstrate the efficacy of our model with various simulations that showcase the different aspects of porous flow behavior.
To summarize, our system of strongly coupled non-pressure forces and enforced incompressibility across overlapping phases allows us to naturally model and stably simulate complex porous interactions.
        \end{abstract}
        \vskip 0.3in
    \end{@twocolumnfalse}
]
\begin{figure*}[!h]
   \begin{center}
   	\includegraphics[width=.245\linewidth,trim={550 0 550 100},clip]{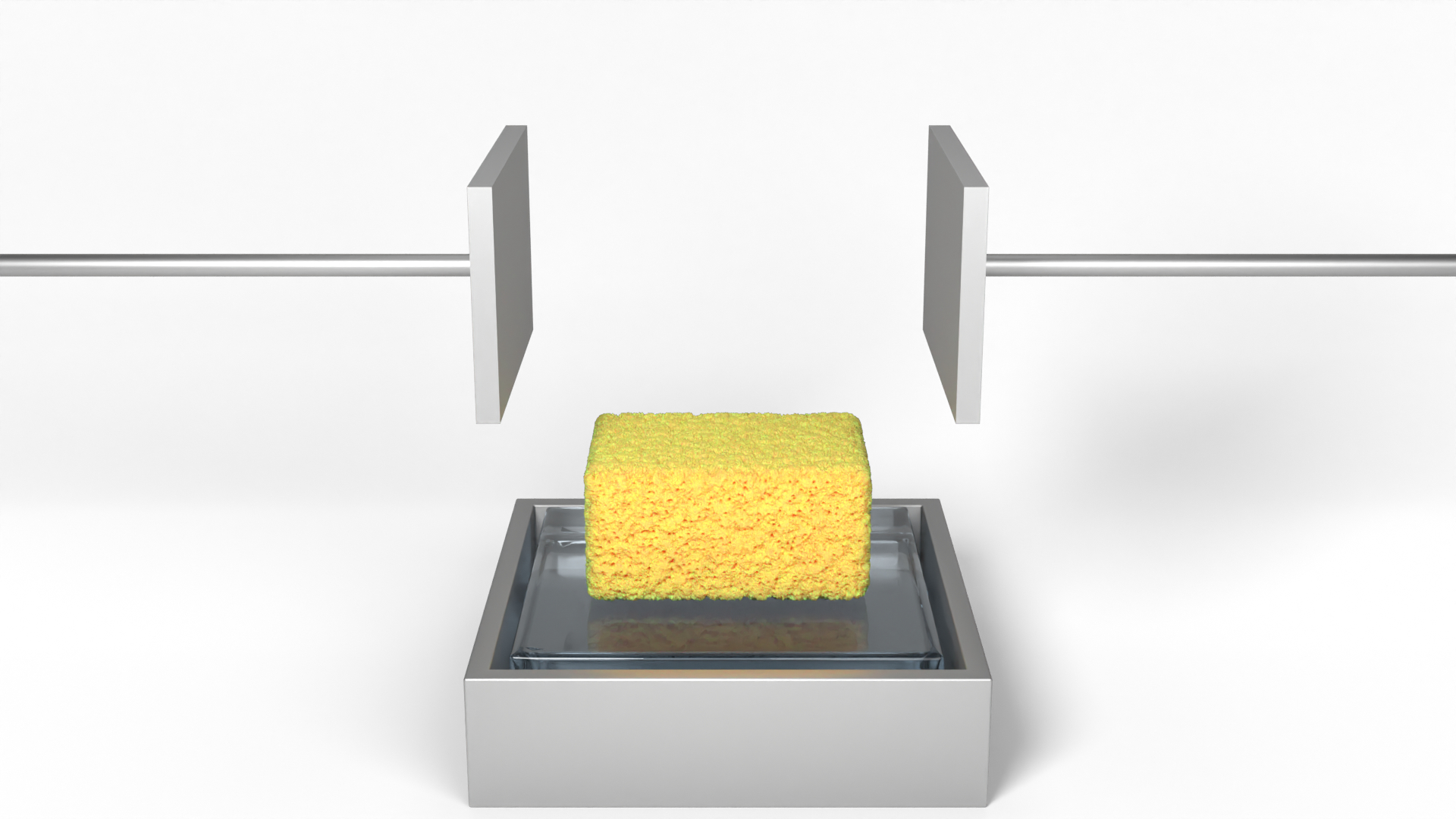}
   	\hfill
   	\includegraphics[width=.245\linewidth,trim={550 0 550 100},clip]{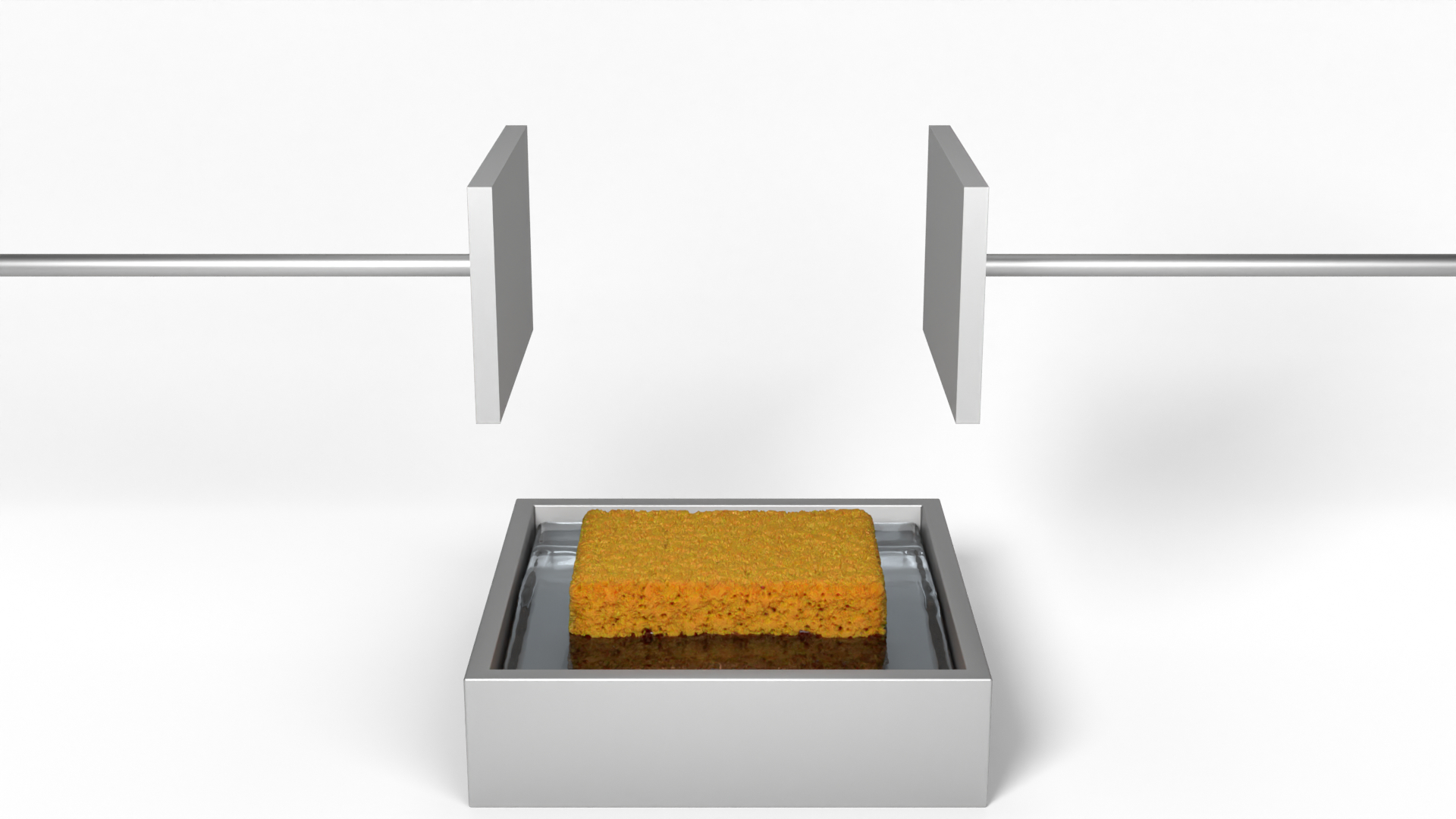}
   	\hfill
   	\includegraphics[width=.245\linewidth,trim={550 0 550 100},clip]{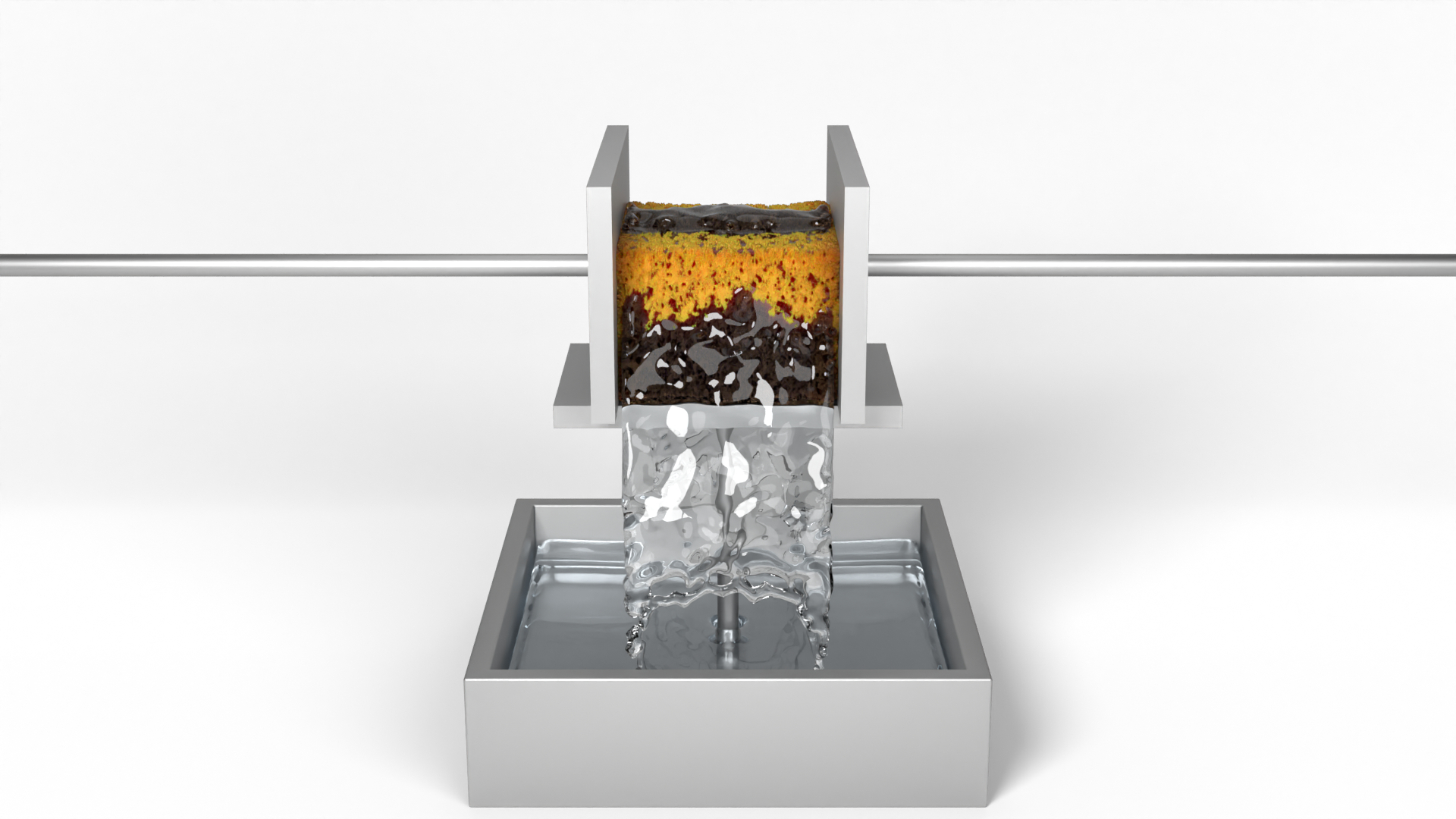}
   	\hfill
   	\includegraphics[width=.245\linewidth,trim={550 0 550 100},clip]{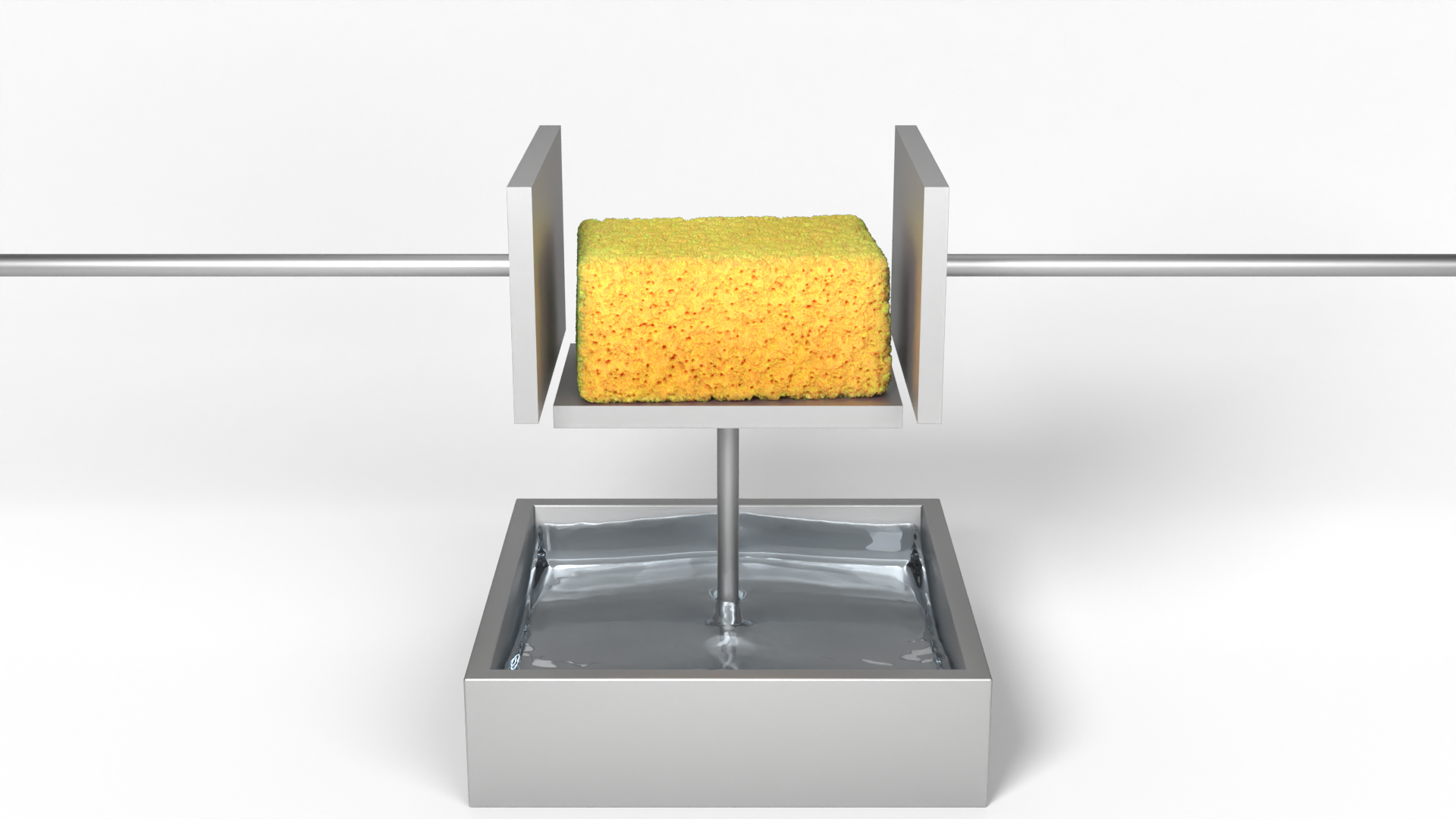}
   \end{center}
   \caption{Simulation of a sponge being soaked in water and subsequently squeezed. Our implicit incompressible porous flow solver is able to model the water leaving the sponge due to pressure forces that consider available pore space. These enable the two phases to overlap in a physically consistent manner, while the porous flow effects are simulated using momentum conserving coupling forces, including capillary action and drag.}
   \label{fig:teaser}
\end{figure*}
% 
% Keywords section (optional)

% > 510.295pt.
% \showthe\textwidth
%                         
% > 243.14749pt.
% \showthe\linewidth

\section{Introduction}

Most natural solids, like wood, soil, or wool, contain holes or small pores, which allow fluid to enter and flow through them.
Even though this porous flow is very common, many simulations that couple solids and fluids ignore the porous aspects.
This greatly reduces realism, since porous flow mechanics can have a huge influence on the macroscopic behavior of both phases.

The Smoothed Particle Hydrodynamics (SPH) method in particular is often used for multiphysics scenarios, since the purely particle-based simulation of all materials readily allows for unified discretization of interaction forces \citep{HJL25}. 
Still, we found the current state of the art to be lacking in regards to coupling solid and fluid phases through porous flow, which would further play into this strength of the SPH method.
We want to simulate complex scenarios, driven by momentum conserving interaction forces between the fluid and solid phase.
Additionally, we need to consider the available pore space, which is vital to determine how much fluid can be absorbed.

Previous SPH porous flow methods in computer graphics model absorption by scaling fluid particles based on how much of their mass is currently outside of the solid domain \citep{Lenaerts08, Ren21}.
Therefore, fluid particles can have vastly different sizes at the solid-fluid interface, that correspond to large mass differences between interacting particles.
This leads to badly conditioned systems in implicit dual solvers \citep{Macklin20} commonly used for SPH pressure computations \citep{Koschier22}.
The fluid movement inside the porous solid is then solved explicitly and simplified through the use of Darcy's law, which describes stationary porous flow based on linear drag forces.
These methods do not ensure balance of forces between the two phases and porous flow effects are modeled using heuristics for the pressure fields.

In this work, we propose a new SPH porous flow method, that does not require scaling of fluid particles and permits the usage of larger time steps by solving an implicit system.
Our method allows fluid and solid particles to overlap by introducing a new density estimation that takes solid porosity into account.
Further, we use an SPH pressure solver to guarantee fluid incompressibility even inside the porous domain.
This allows us to model other porous flow effects with momentum conserving coupling forces, that are motivated by the physical forces acting between the fluid and solid, instead of using heuristics.
Here, we take inspiration from granular flow models using the material point method (MPM) \citep{Tampubolon17, Gao18}, but derive new implicit interaction forces using state-of-the-art SPH solvers for effects like viscosity and surface tension.
This allows us to present the, to the best of our knowledge, first implicit porous flow solver using SPH in computer graphics.
For coupling with deformable solids, which we simulate using corotated linear elasticity, we propose modifications to include softening and bloating induced by absorbed fluid content and derive a strongly coupled implicit solver, which can handle large drag forces between solid and fluid phase.
Finally, we propose a new model for capillary action using adhesion forces and include buoyancy based on fluid pressure, such that our method is able to simulate a wide range of porous flow effects (see \fig{fig:teaser}).

In summary, our contributions are:
\begin{itemize}
	\item the derivation of a physical model for porous flow, that includes drag, capillary action and buoyancy,
	\item an SPH framework that includes a pressure solver for overlapping particles in porous flow regions,
	\item modifications to linear stress computation to include porous solid effects, like softening and bloating,
	\item and a strongly coupled implicit linear formulation of the non-pressure forces.
\end{itemize}

 \section{Related Work}
	
Our work investigates  the macroscopic effects of the microscopic phenomenon of fluid moving within porous solids.
We model this behavior using a continuum model, which is discretized using SPH.
Based on the specific material properties, different approaches to model porous flow have already been developed.
In this section, we will discuss related literature regarding general continuum porous-flow simulation using SPH-based methods in particular.

\paragraph{Smoothed Particle Hydrodynamics}

The SPH method, being a truly mesh-free particle-based Lagrangian method, is well suited to model phenomena with complex and changing interfaces, such as free-surface or porous flow.
In recent years, this method has seen many improvements regarding stability, efficiency, and the ability to model a wide range of physical effects.
These include, among many others, fluid viscosity \citep{PICT15,Weiler18}, elasticity \citep{Peer17,KBF+21} and surface tension \citep{AAT13,ZRS20,Jeske23}.
These effects greatly influence porous phenomena, such as viscous drag or capillary action due to adhesion forces between the fluid and the solid phase.
We therefore see great potential in using SPH to simulate porous flow and build on these models to derive strongly coupled interaction forces between the two phases.

To ensure that the fluid can only fill the available pore space, we use an SPH pressure solver to limit local fluid density.
Modern pressure solvers ensure incompressibility by solving an implicit pressure Poisson equation \cite{KBST19}.
\citet{Solenthaler08} additionally show how density estimation can be adapted for the computation of pressure forces between non-dilutable fluid mixtures.
We instead propose a new density estimation for overlapping particles based on porosity, which ensures fluid incompressibility while also prohibiting oversaturation of pores.

\paragraph{Porous Flow Simulation}

Inspired by \citet{Lenaerts08}, many porous flow models in computer graphics use the SPH method for the non-absorbed part of the fluid phase, while the absorbed fluid mass is stored as a virtual saturation or wetness term on the solid discretization elements.
In these methods, the number and size of fluid particles grows and shrinks to reflect how much fluid is currently outside of the porous object.
Implementing this method is therefore challenging, as inserting particles in areas already filled with fluid is not a straightforward process and can lead to discontinuities \citep{Winchenbach21}.
The absorbed fluid on the other hand is transported between the solid elements based on diffusion equations.
This approach can therefore be used with various solid simulation methods, like SPH \citep{Lenaerts08, Lenaerts09}, the discrete element method (DEM) \citep{Rungjiratananon08}, discrete elastic rods (DER) \citep{Lin15} or mesh-based methods \citep{Huber11, Patkar13}.

Other approaches use multiphase methods, where each particle represents both solid and fluid at once \citep{Yan16, Ma22}.
The exact local mixture is then tracked using volume fractions stored at the particles.
These approaches are well suited to simulate dissolution processes, but the vastly different behavior of the two phases can make particle movement not well defined.

For other porous flow applications, an overlapping domain approach has therefore become more popular.
Here, both phases are simulated based on their own constitutive models and respective discretization elements, which are allowed to occupy the same space as the other phase.
This approach allows the use of specialized models for each phase and can make use of the inherent mass preserving property of particle-based simulation methods, as no mass transfer between discretization elements is required.
\citet{Ren21} recently proposed a new SPH porous flow model that uses this kind of approach.
While their method allows the coupling of porous materials with multiphase fluids, fluid particles entering the porous region is again made possible by reducing their size during the absorption process.
\citet{Ren21} employ harmonic means for volume terms in the pressure solver to increase stability, but we still encountered issues in scenarios with fast porous flow where particles can rapidly change in size.
In contrast to our approach, their method does not correct densities of absorbed fluid particles and only propagates them using Darcy's law, using pore pressure heuristics in a way that does not consider momentum conservation between fluid and solid.

Other methods that allow particle overlapping instead simulate fluid transport through the porous domain using momentum-conserving interaction forces.
This approach is mostly used for the mixture of granular materials and fluids \citep{Wang21, Bui17}.
Our method follows this idea for general porous flow scenarios, while both phases are simulated using SPH.

We adapt the IISPH pressure solver \citep{Ihmsen14} to allow overlapping particles, while still taking the presence of the other material into account.
To the best of our knowledge, this is a novel contribution, as other porous flow approaches that use SPH do not account for the available pore space in the overlapping regions \citep{Ren21} or only use explicit pressure forces \citep{Bui17}, that require small time step sizes.

Aside from SPH, particle-grid hybrid methods like MPM \citep{Jiang16} are a popular alternative to simulate porous effects.
MPM methods have been used to great effect to simulate porous materials, such as fluid interacting with hair \citep{Fei19}, cloth \citep{Fei18}, or granular materials \citep{Tampubolon17, Gao18}, just to mention a few.
However, our goal is to enable physically consistent porous flow simulation within SPH frameworks, where MPM approaches cannot be directly applied due to the different discretization techniques.
\section{Method}

\begin{figure}
	\includegraphics[width=.85\linewidth]{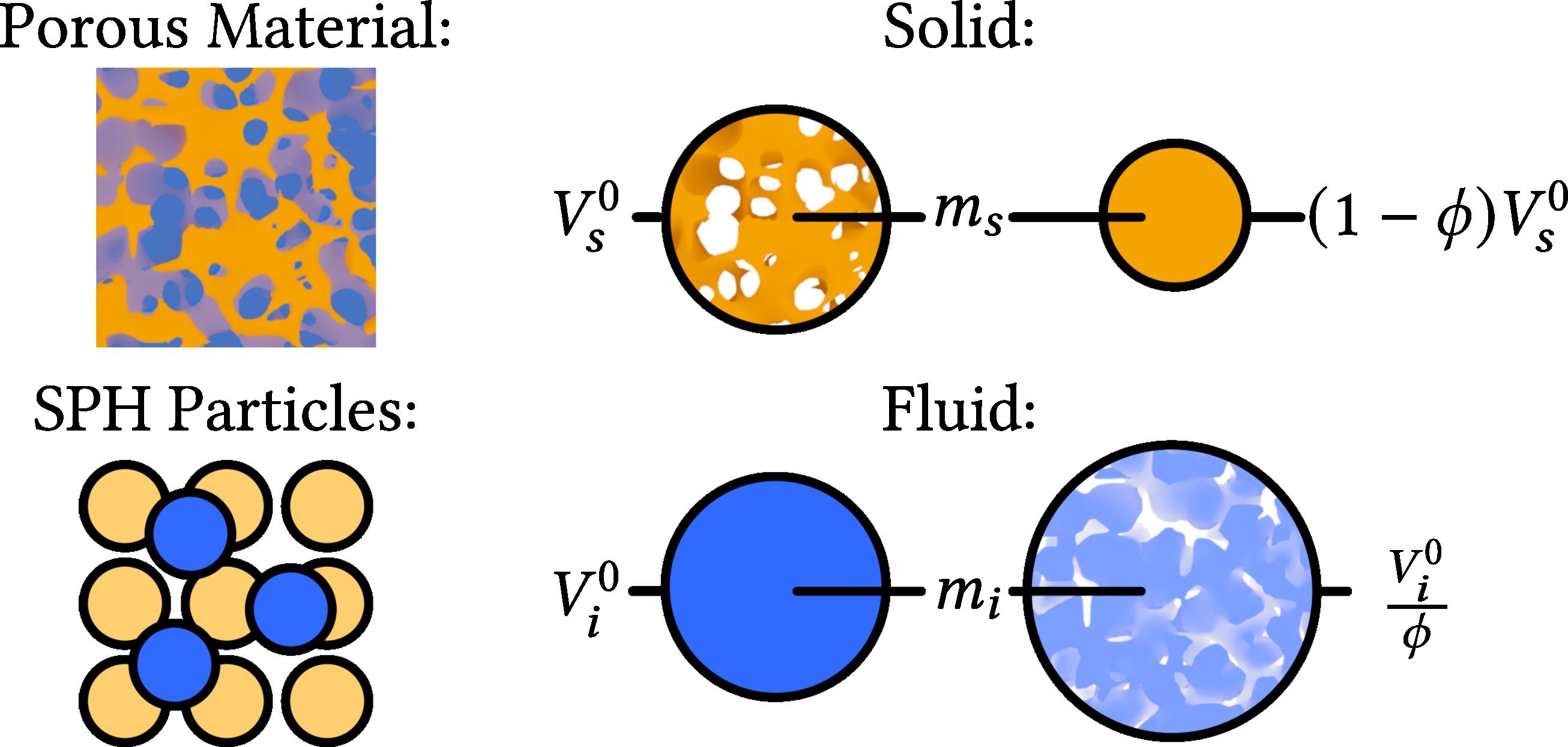}
	\caption{Particle volume definitions for both solid and fluid phase. A solid particle $\spID$ represents the porous object, such that its sampling volume $\spexp{V}{0}$ includes void space based on the porosity $\phi$. The same solid mass $\spp{m}$ without pores only covers the volume $(1 - \phi) \spexp{V}{0}$.
		The volume $V_i^0$ of fluid particle $\fpID$ on the other hand only represents the fluid phase, causing the particles to spread out when they overlap with the solid.}
	\label{fig:volume-averaging}
\end{figure}

If the internal structure of a solid contains a connected network of pores, fluid from the outside can freely enter into and move through the resulting open space, giving rise to the porous flow phenomenon.
Here, the very small scale of the pores results in macroscopic behavior not usually seen in other fluid simulation contexts.
One well-known example is capillary rise, caused by surface tension forces, where fluid is pulled upwards against gravity.
While porous flow properties often depend on the specific internal solid structure or fluid properties, we did not tune our model to specific materials.
We instead provide a general porous flow framework that includes the most common associated effects and results in visually intriguing simulations with great artistic control.

In our work, we simulate porous flow from a macroscopic perspective, since it would be vastly inefficient to resolve all the microscopic pore details. 
We therefore adopt a continuum approach and account for the pore volume inside of a solid just by using the porosity value $\phi \in [0,1)$.
This value is a measure for the free space inside the porous object, such that a non-porous material has a porosity of $\phi = 0$, while a porosity of $\phi = 1$ corresponds to pure void space.
Conceptually, this can be seen in \fig{fig:volume-averaging}, where a solid particle $\spID$ with mass $\spp{m}$ has a volume of $\spexp{V}{0}$, which is only partially occupied by the solid phase itself.

We use SPH to simulate both fluid and solid, where each particle represents a material parcel from one of the two phases.
These two sets of particles behave according to their own equations of motion, but interact with particles of the other set through momentum conserving interaction forces.
In this section, we first derive a new way to estimate particle densities for porous mixtures, which accounts for the pore space in regions where solid and fluid particles overlap.
Based on these density estimations, we propose formulations for pressure accelerations that counteract compression and ensure that fluid absorption is limited by the porosity.

Next, we introduce implicit formulations for the different forces required to solve the equations of motion, to which we add solid-fluid coupling forces for drag, buoyancy and capillary action.
These forces are assembled in a linear system for the particle velocities, that strongly couples the interaction forces with other non-pressure forces, like elastic deformation response and fluid viscosity.

\subsection{Porous Densities} \label{sec:densities}

In the SPH method, a continuous material is sampled with particles, where each particle corresponds to a parcel with constant mass.
Given the sampling volume $V_i^0$ and the constant rest density $\rho_i^0$ of the material, the mass $m_i$ of particle $i$ is $m_i  = \rho_i^0 V_i^0.$
The current density $\rho_i$ is approximated using the mass of all nearby particles $j$ of the same phase:
\begin{equation}
	\label{eqn:SPH_density_gen}
	\rho_i = \sum_j m_j W_{ij},
\end{equation}
where $W_{ij} = W(\v{x}_i - \v{x}_j, h)$ is a normalized kernel with smoothing length $h$, that assigns a weight based on the distance between the particle positions $\v{x}_i$ and $\v{x}_j$.
This equation directly links particle spacing to density and $\rho_i \approx \rho_i^0$ if each particle covers a volume that is equal to its sampling volume $V_i^0$, in which case $\sum_j V_j^0 W_{ij} \approx 1$ holds.
We refer to the tutorial by \citet{KBST19} for the derivation and more information on SPH.

The density estimation plays a critical role in pressure solvers, which ensure that the distances between particles correspond to their physically accurate volume.
While \eq{eqn:SPH_density_gen} gives a decent estimate if only one phase occupies the region around particle $i$, in the case of porous flow we also need to consider the presence of the other phase.
We therefore introduce a new density estimation to account for porosity, based on our solid sampling.
To differentiate particles, we will use the following indices based on their phase:
fluid particles $\fpID$,
fluid neighbors $\fnID$,
solid particles $\spID$, and
solid neighbors $\snID$.

\subsubsection{Porous Solid Density}

We assume that the fluid phase and the solid phase itself (but not necessarily the pores) are incompressible.
Both phases then have to fulfill the continuity equation, which links the material derivative of the density $\frac{D \rho}{D t}$ to the velocity field $\v{v}$:
\begin{equation}
	\label{eqn:conti}
	\frac{D \rho}{D t} = -\rho \nabla \cdot \v{v} \equiv 0.
\end{equation}

Since we want to model porous behavior from a macroscopic perspective, we consider porous solids as continuous materials that include both the solid phase and the so-called void regions that can be filled with a fluid phase.
We further differentiate between two types: those that allow compression of pores and those that do not.
The first group includes materials like sponges, which can be compressed until all pores have collapsed, while the solid phase itself is incompressible.
The second group includes materials like wood or compact soil, which are resistant to any compression due to the structural integrity of the solid skeleton.
In this section, we derive an estimation for the density of the porous object, while the different behavior of the two groups is achieved using SPH pressure solver properties, as detailed in Sec.~\ref{sec:pressure-solver}.

For the SPH discretization, we sample porous objects in their rest configuration using particles $s$ with volumes $\spexp{V}{0}$ (see \fig{fig:volume-averaging}).
Given the constant rest density of the solid phase $\spexp{\rho}{0}$, the particle $\spID$ has an effective rest density of $(1-\phi) \spexp{\rho}{0} \leq \spexp{\rho}{0}$ due to the constant rest porosity $\phi$.
We therefore define the solid particle mass as
	$\spp{m}  = (1 - \phi) \spexp{\rho}{0} \spexp{V}{0}.$
Inserting this particle mass definition into \eq{eqn:SPH_density_gen} and summing only over solid neighbors $t \in \mathcal{N}^{S}$ results in:
\begin{equation}
	\label{eqn:SPH_density_solid}
	\begin{aligned}
		\spp{\rho} &= (1 - \phi) \spexp{\rho}{0} \sum_{\insolidneighborhood{\spID}} \snexp{V}{0} \Wss,
	\end{aligned}
\end{equation}
where we exploit that the solid has a constant rest density ($\spexp{\rho}{0} = \snexp{\rho}{0}$).
Due to our sampling choice, $\spp{\rho}$ approximates the density $(1 - \phi) \spexp{\rho}{0}$ of the porous object and not that of the solid phase itself.
Therefore, $\spp{\rho}$ should be constant for incompressible porous objects.
In the case of compressible porous objects, the pores can be squeezed out such that densities greater than $(1 - \phi) \spexp{\rho}{0}$ are allowed.
This corresponds to solid particles moving closer together, as \eq{eqn:SPH_density_solid} increases for denser particle configurations.
In this case, the particle's estimated density $\spp{\rho}$ should never exceed the solid phase density $\spexp{\rho}{0}$, at which point the porous object does not contain void space anymore.
Further volume loss corresponds to the solid phase itself being compressed, which we prohibit in our method.
These density constraints can be simulated using SPH pressure solvers, as will be discussed later.

\subsubsection{Fluid Density}

\begin{figure}
	\begin{subfigure}{.32\linewidth}
		\includegraphics[width=\linewidth,trim={0 0 0 0},clip]{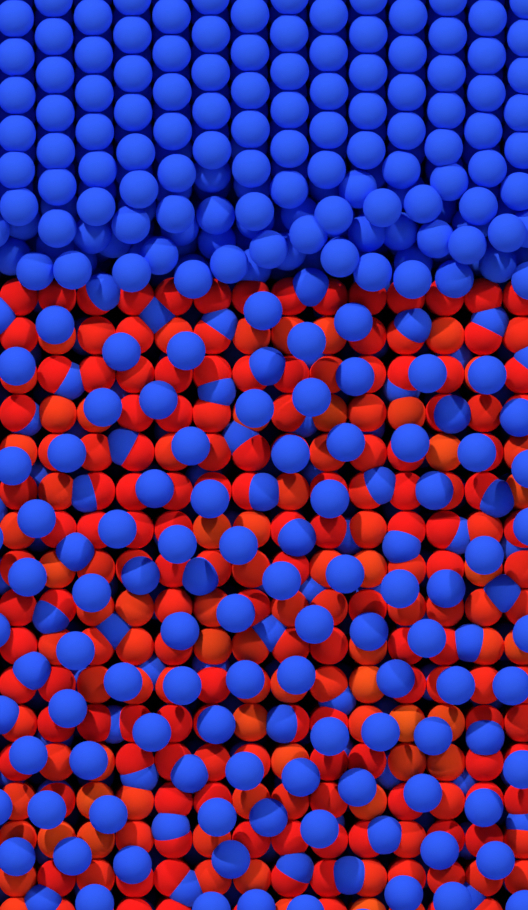}
		\caption{$\phi = 0.4$}
	\end{subfigure}
	\begin{subfigure}{.32\linewidth}
		\includegraphics[width=\linewidth,trim={0 0 0 0},clip]{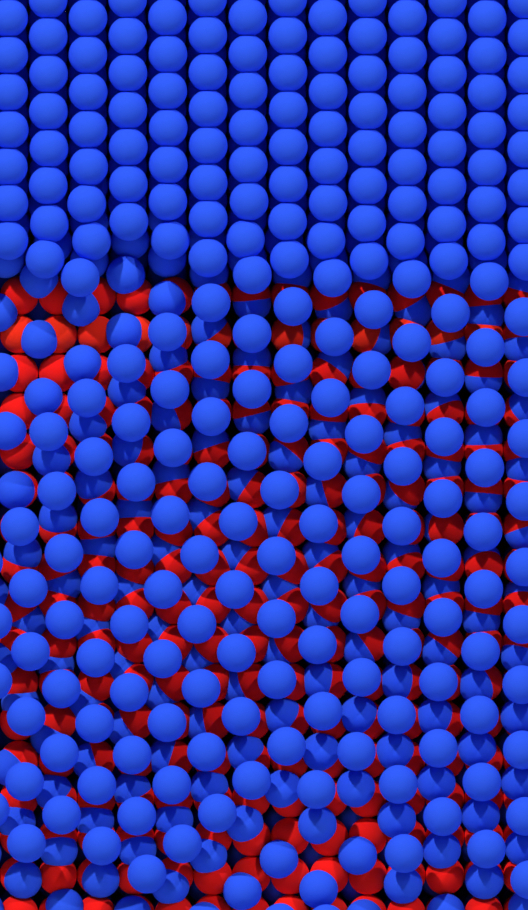}
		\caption{$\phi = 0.6$}
	\end{subfigure}
	\begin{subfigure}{.32\linewidth}
		\includegraphics[width=\linewidth,trim={0 0 0 0},clip]{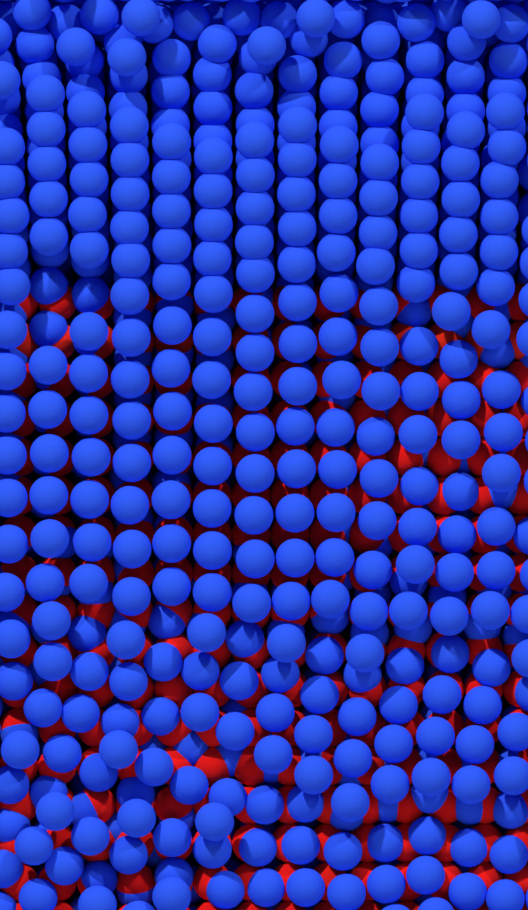}
		\caption{$\phi = 0.8$}
	\end{subfigure}
	\caption{Fluid seeping into a porous object, where an internal view of the particle distributions is achieved using cut planes. Shown is a closeup of the saturated states around the solid-fluid interface for different solid porosities. The porosity determines the free pore space inside the solid and therefore the distance between fluid particles.}
	\label{fig:porosity}
\end{figure}

We sample the continuous fluid phase with particles $\fpID$, where we assume that the sampling volume $\fpexp{V}{0}$ is completely filled by the fluid phase.
This is different from our definition of the porous solid phase, as shown in \fig{fig:volume-averaging}.
Outside of the porous domain, we can use \eq{eqn:SPH_density_gen} to estimate the fluid particle densities $\fpp{\rho}$.
However, when the fluid particle overlaps with the solid particles we need to account for the space already occupied by the solid phase.

We take inspiration from the number density approach \citep{Solenthaler08}, where all neighbors of particle $i$ are included in the density estimation while all particles are treated as having the same phase as $i$.
This approach allows for consistent pressure coupling between different phases but prohibits particle overlapping.
For porous flow, we instead want fluid particles to be able to occupy the same space as the solid particles, based on their porosity.
We achieve this by only including the part that corresponds to the solid phase itself (without pores) in the density estimation of fluid particles.
Since the porosity $\phi$ measures pore space inside the porous solid, we can exclude this space from the solid particle sampling volumes $\spexp{V}{0}$ by multiplying with the factor $1-\phi$.
We then approximate the density of fluid particle $\fpID$ as:
\begin{equation}
	\label{eqn:SPH_density_covered}
	\fpp{\hat{\rho}} = \fpexp{\rho}{0} \sum_{\influidneighborhood{\fpID}} \fnexp{V}{0} \Wff + \fpexp{\rho}{0} \sum_{\insolidneighborhood{\fpID}} (1 - \phi) \snexp{V}{0} \Wfs.
\end{equation}
This results in a term for the solid particles that is similar to our proposed solid density estimation (\eq{eqn:SPH_density_solid}), except that the sum approximating the local solid volume fraction is multiplied with the fluid rest density instead of the solid one.
If no solid particles are close to the fluid particle, the term for the solid particles vanishes.
In that case, \eq{eqn:SPH_density_covered} only contains terms from the fluid neighborhood $\mathcal{N}^F$, and we recover \eq{eqn:SPH_density_gen}.
Therefore, $\fpp{\hat{\rho}}$ is consistent both outside and inside the porous domain.
The porosity $\phi$ now limits how much fluid is allowed to overlap with the solid.
This also influences fluid particle spacing, as shown in \fig{fig:porosity}.

\subsection{Pressure Solver} \label{sec:pressure-solver}
We enforce incompressibility by solving a pressure Poisson equation:
\begin{equation}
	\label{eqn:PPE}
	\dt \nabla^2 p = \frac{\rho^0 - \rho^*}{\dt},
\end{equation}
where $p$ is the pressure, $\rho^*$ is the predicted density based on the non-pressure forces and $\rho^0$ is the rest density of the material.

As shown by \citet{Koschier22}, most implicit SPH pressure solvers use an equation like \eq{eqn:PPE} to counteract compression.
These solvers require the density change based on velocities, to be able to predict $\rho^*$, and a formulation for the accelerations resulting from pressure gradients.
The former we derive by taking the temporal derivative of \eq{eqn:SPH_density_solid} and \eq{eqn:SPH_density_covered}, where $\v{v}_{\fpID \fnID} = \v{v}_{\fpID}  -\v{v}_{\fnID}$:
\begin{equation}
	\label{eqn:SPH_density_change}
	\begin{aligned}
	\frac{D \spp{\rho}}{D t} = &-(1 - \phi)\spexp{\rho}{0}  \sum_{\insolidneighborhood{\spID}} \snexp{V}{0} \v{v}_{\snID \spID} \cdot \nabla \Wss, \\
	\frac{D \fpp{\hat{\rho}}}{D t} = &-\fpexp{\rho}{0} \sum_{\influidneighborhood{\fpID}} \fpexp{V}{0} \v{v}_{\fnID \fpID} \cdot \nabla \Wff \\
	&- \fpexp{\rho}{0} \sum_{\insolidneighborhood{\fpID}} (1 - \phi) \snexp{V}{0} \v{v}_{\snID \fpID} \cdot \nabla \Wfs.
	\end{aligned}
\end{equation}
Note that these equations are SPH approximations of the velocity divergence, which is consistent with \eq{eqn:conti}.

The pressure accelerations on the other hand result from the pressure gradient.
For fluid particles, we treat the solid particles similar to boundary particles and adapt the method presented by \citet{BWJ23} to our porous density estimation, with corrected sampling volumes as in \eq{eqn:SPH_density_covered}:
\begin{equation}
	\label{eqn:pressure_accel_fluid}
	\begin{aligned}
		\fpexp{\v{a}}{\text{press}} = &- \fpexp{\rho}{0} \sum_{\influidneighborhood{\fpID}} \fnexp{V}{0} \left( \frac{\fpp{p}}{\fpexp{\rho}{2}} + \frac{\fn{p}}{\fnexp{\rho}{2}} \right) \nabla \Wff \\
		&- \fpexp{\rho}{0} \sum_{\insolidneighborhood{\fpID}} (1 - \phi) \snexp{V}{0} \frac{\fpp{p}}{\fpexp{\rho}{2}}  \nabla \Wfs.
	\end{aligned}
\end{equation}
Here, we chose the symmetric formula for the gradient estimation, which is commonly used in SPH pressure solvers. This formulation produces pressure forces that encourage ordered particle patterns, that in turn enhance the accuracy of all other SPH approximations \citep{Price12}.
For solid particles, we do not consider any pressure forces acting from nearby fluid particles at this stage and instead use:
\begin{equation}
	\label{eqn:pressure_accel_solid}
	\spexp{\v{a}}{\text{press}} = -  (1 - \phi)\spexp{\rho}{0} \sum_{\insolidneighborhood{\spID}} \snexp{V}{0} \left( \frac{\spp{p}}{\spexp{\rho}{2}} + \frac{\sn{p}}{\snexp{\rho}{2}} \right) \nabla \Wss.
\end{equation}
The interaction with fluid particles is handled separately as an additional coupling force to simulate buoyancy.
Since elastic forces also counteract volume change, which is not considered in the pressure solver, we achieve more plausible results by coupling this interaction force directly with the elasticity solver.

In SPH, it is common to use pressure clamping \citep{KBST19}, such that only compression is considered as a violation of the continuity equation.
Hence, the right-hand side of \eq{eqn:PPE} is required to be negative for pressure forces to be active.
For solid particles, we can use this property to switch between compressible and incompressible porous objects.
In the first case, all densities $\spp{\rho}$ smaller than $\spexp{\rho}{0}$ are acceptable.
For incompressible porous materials, we instead only allow densities under $(1 - \phi) \spexp{\rho}{0}$, to reflect that we do not allow reduction of the pore space.
This is achieved by replacing the rest density term in \eq{eqn:PPE}.
For fluid particles, we always require that their density is smaller or equal to the rest density $\fpexp{\rho}{0}$.
More details on this and the derivation of the density constraints are given in the supplementary document.

We used a modified IISPH pressure solver \citep{Ihmsen14} to solve \eq{eqn:PPE}, using the updated density change and pressure acceleration terms.
This pressure solver uses the predicted density changes to compute pressure values for each particle, such that the resulting accelerations remove any density deviations.
The resulting implicit linear system is solved using the relaxed Jacobi method, where in each iteration the pressure values are clamped to be positive, such that they only counteract compression.
For boundary handling we use volume maps \citep{Bender19a} or the approach by \citet{Akinci12}, but our method does not have strict requirements in this regard.

\subsection{Porous Interaction Forces} \label{sec:porous-forces}

In the previous section, we presented how we can allow fluid and solid particles to overlap, while still adhering to the continuity equation.
The pressure forces do not depend on any specific pore structure or fluid properties, since they directly follow from incompressibility and the available pore space given by the porosity value.
Next, we need to cover all other forces that act on the fluid and solid particles according to their equations of motion.
We start with the porous interaction forces, which are less well defined and greatly depend on the material properties.
Here, we chose force formulations that not only cover most of the common porous flow effects but can also be combined into a unified implicit solver.

For a solid particle $\spID$, we use a modified Cauchy momentum equation, which includes external forces $\spexp{\v{f}}{\text{ext}}$ and the porous coupling forces $\spexp{\v{f}}{\text{pore}}$:
\begin{equation}
	\label{eqn:solid_momentum}
	\frac{D \spp{\v{v}}}{D t} = \frac{1}{ \spp{\rho}} \nabla \cdot \boldsymbol{\sigma} + \frac{1}{\spp{m}} \spexp{\v{f}}{\text{pore}} + \frac{1}{\spp{m}} \spexp{\v{f}}{\text{ext}},
\end{equation}
where $\boldsymbol{\sigma}$ is the local stress tensor.
For a fluid particle $\fpID$, we instead add the porous forces to the Navier-Stokes momentum equation:
\begin{equation}
	\label{eqn:navier_stokes}
	\frac{D \fpp{\v{v}}}{D t} = -\frac{1}{\fpp{\hat{\rho}}} \nabla p + \frac{\fpp{\mu}}{\fpp{\hat{\rho}}} \nabla^2 \v{v} +  \frac{1}{\fpp{m}} \fpexp{\v{f}}{\text{pore}} + \frac{1}{\fpp{m}} \fpexp{\v{f}}{\text{ext}}.
\end{equation}
For our method, we formulate $\v{f}^{\text{pore}}$ as the sum of the most common effects observed in porous flow, which are capillary action $\v{f}^{\text{cap}}$, drag $\v{f}^{\text{drag}}$, and buoyancy $\v{f}^{\text{buo}}$.
Often, the drag forces inside the porous solid are very strong, which makes explicit solvers highly unstable.
Instead, we propose implicit formulations for all non-pressure forces, that depend linearly on the velocities $\v{v}^{n+1}$ of the next time step $n+1$ and can easily be combined into one strongly coupled system.

\subsubsection{Capillary Action}

Capillary forces allow fluid to rise inside the porous material, due to adhesive forces acting between the solid and the fluid. 
To model this effect, we simplify the adhesion model by \citet{Jeske23} and formulate attraction force between fluid and solid using a capillary coefficient $C^{\text{cap}}$ as:
	\begin{equation}
		\label{eqn:adhesion}
		\fpexp{\v{f}}{\text{cap}} = -\sum_{\insolidneighborhood{\fpID}} C^{\text{cap}}(\snexp{S}{n}) \frac{\bar{m}_{\fpID \snID}}{\bar{\rho}_{\fpID \snID}^{n}} \v{x}_{\fpID \snID}^{n+1} W_{\fpID \snID}^n,
	\end{equation}
where $\bar{m}_{ij} = \frac{m_i + m_j}{2}$ and $\bar{\rho}_{it} = \frac{\rho_i + (1 - \phi) \rho_t^0}{2}$ are averaged to ensure momentum balance, $S$ is the saturation of the solid particle and $\v{x}_{i t} = \v{x}_{i} - \v{x}_{t}$.
For a constant $C^{\text{cap}}$, this formulation yields a vector which points into the direction with the most solid neighbors, such that the resulting acceleration will pull fluid particles inside the porous body.
To include increased attraction forces towards unsaturated regions we further weight them based on solid particle saturation.

\begin{figure}
	\begin{subfigure}{.5\linewidth}
		\centering
		\includegraphics[width=.99\linewidth,trim={400 200 957 250},clip]{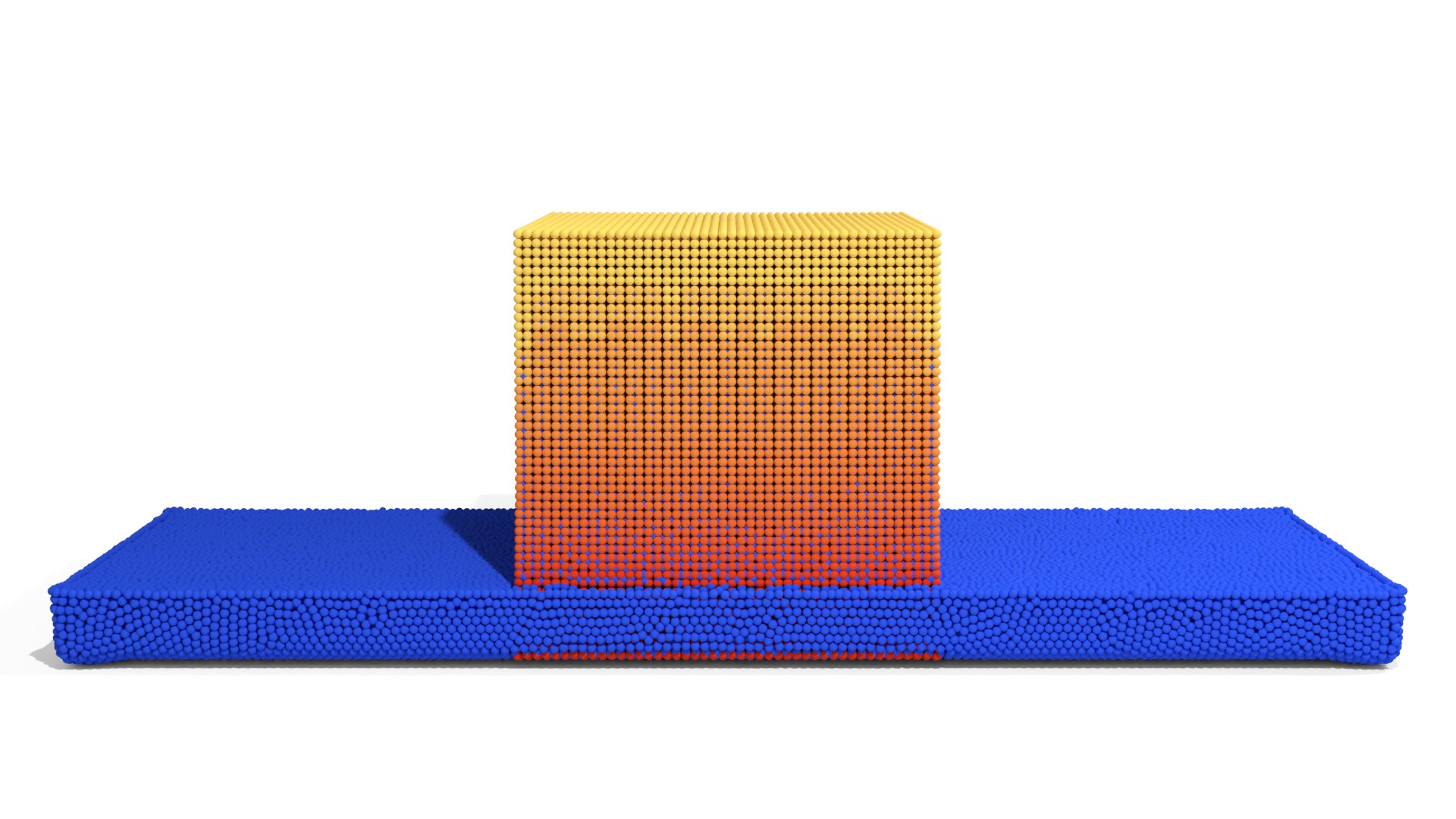}
		\caption{$\eta_{\text{cap}} = 1.0$}
	\end{subfigure}%
	\begin{subfigure}{.5\linewidth}
		\centering
		\includegraphics[width=.99\linewidth,trim={957 200 400 250},clip]{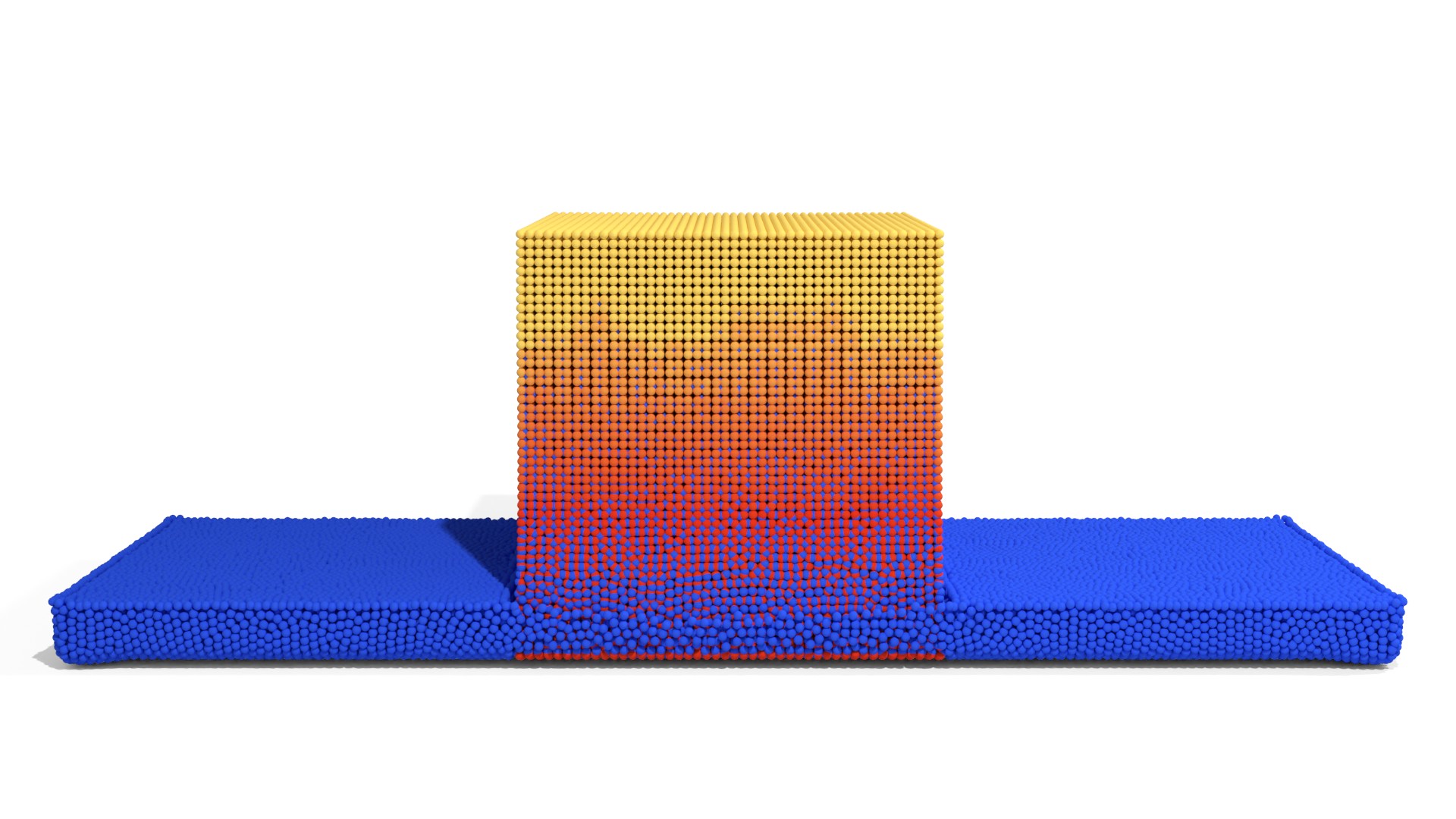}
		\caption{$\eta_{\text{cap}} = 0.5$}
	\end{subfigure}
	\caption{Porous block absorbing water after 20\si{\second} with $C^{\text{cap},0} = 500 \si{\newton \per \meter}$ and two different capillary potential falloff parameters. The solid particles are colored based on saturation ranging from yellow (dry) to red (fully saturated). In (a), fully saturated solid particles no longer produce an adhesion force. A reduced falloff, as shown in (b), allows for more fluid to be pulled into the lower region of the block, leading to more total fluid absorption and a higher fully saturated region.}
	\label{fig:adhesion_falloff}
\end{figure}

The saturation $\spp{S}$ measures how much the local pores are already filled with fluid, which we estimate by comparing the fraction of the local fluid and solid volume with the porosity:
\begin{equation}
	\label{eqn:saturation}
	\spp{S} = \frac{1}{\phi} \frac{\sum_{\influidneighborhood{\spID}} \fnexp{V}{0} \Wsf}{\sum_{\insolidneighborhood{\spID}} \frac{\sn{m}}{\sn{\rho}} \Wss}.
\end{equation}
To avoid issues arising from interpolation errors at the solid-fluid interface, we clamp saturation to ensure $\spp{S} \in [0,1]$.
We then decrease the capillary potential for increasing saturation, based on a falloff factor $\eta_{\text{cap}} \in [0,1]$:
\begin{equation}
	\label{eqn:adhesion_mod}
	C^{\text{cap}}(\spp{S}) = C^{\text{cap},0} (1 - \spp{S} \eta_{\text{cap}}),
\end{equation}
where the adhesion $C^{\text{cap},0}$ determines the maximum height fluid can reach based on capillary action.

As capillary pressure depends on the channel radius, smaller pores can pull fluid much higher than the larger pores can, such that only a portion of the solid will be fully saturated \citep{Ha18}.
Our model allows direct control over the final height of this fully saturated region, based on the falloff factor $\eta_{\text{cap}}$.
As shown in \fig{fig:adhesion_falloff}, this front will move upwards and closer to the maximum height when decreasing $\eta_{\text{cap}}$.
\citet{Lenaerts08} also propose to decrease capillary potential based on saturation, but use a power function which always reduces the adhesion force to zero for fully saturated regions.
For $\eta_{\text{cap}} < 1$, we instead include adhesion forces for saturated solid particles, which then still attract fluid particles that are close to the porous solid boundary.
This absorption force in turn results in pressure forces acting on already absorbed fluid, which is pushed further inside towards regions with available pore space, if there are no opposing forces stronger than the adhesion.
This allows gravitational, pressure and capillary forces to balance out in each time step, resulting in stable configurations even when the fully saturated region lies above the outer fluid surface.
	
For the averaged density $\bar{\rho}_{it}$ we use $\fpp{\rho}$ instead of $\fpp{\hat{\rho}}$ for the fluid particle density.
Since the sum in \eq{eqn:adhesion} approximates an integral over the whole volume around the fluid particle, not just the integral over a single phase, we require the approximated particle influence volumes $\frac{m}{\rho}$ to reflect this.
We therefore use the densities averaged over the whole surrounding volume, not just the volume occupied by the phase.

While the solid particle densities $\spp{\rho}$ already include the pore volumes, we could directly use them in the force estimation. But for large adhesion forces, we observed simulation artifacts when using the approximated solid density.
Since solid particles at the surface of the porous object have a smaller estimated density (due to missing neighbors), their volume is greatly overestimated and fluid particles close to but still outside the porous domain experience unnaturally strong capillary forces.
We instead use the rest density $(1 - \phi) \rho_s^0$ here to solve this problem.

The forces acting on the porous solid follow from momentum balance, since $\v{f}^{\text{cap}}_{\fpID \leftarrow \spID} = - \v{f}^{\text{cap}}_{\spID \leftarrow \fpID}$ must hold:
\begin{equation}
	\label{eqn:adhesion_mirror}
	\spexp{\v{f}}{\text{cap}} = -\sum_{\influidneighborhood{\spID}} C^{\text{cap}} (\spexp{S}{n}) \frac{\bar{m}_{\spID \fnID}}{\bar{\rho}_{\spID \fnID}^{n}} \v{x}_{\spID \fnID}^{n+1} W_{\spID \fnID}^n.
\end{equation}
This formulation of capillary adhesion does not directly include the velocities, but the particle positions. 
Since we aim for formulations linear in $\v{v}^{n+1}$, we approximate $\v{x}^{n+1} = \v{x}^{n} + \Delta t \v{v}^{n+1}$ and split the force into two terms $\mathbf{f} = \mathbf{f}(\mathbf{v}^{n+1}) + \mathbf{b}^n$, e.g. for fluid particles:
\begin{equation}
	\label{eqn:adhesion_semi_implicit}
	\begin{aligned}
		\fpexp{\v{f}}{\text{cap}} \left(\v{v}^{n+1} \right) &= - \Delta t \sum_{\insolidneighborhood{\fpID}} C^{\text{cap}}(\snexp{S}{n}) \frac{\bar{m}_{\fpID \snID}}{\bar{\rho}_{\fpID \snID}^n} \v{v}_{\fpID \snID}^{n+1} W_{\fpID \snID}^n, \\
		\fpexp{\v{b}}{\text{cap},n} &= - \sum_{\insolidneighborhood{\fpID}} C^{\text{cap}}(\snexp{S}{n}) \frac{\bar{m}_{\fpID \snID}}{\bar{\rho}_{\fpID \snID}^n} \v{x}_{\fpID \snID}^{n} W_{\fpID \snID}^n.
	\end{aligned}
\end{equation}
The term $\mathbf{b}^{\text{cap},n}$ only depends on known values and will contribute to the right-hand side of the implicit system.

\subsubsection{Drag}

As proposed by \citet{Bui17}, the drag forces between the fluid and solid particles can be simulated as viscous friction.
In our work, we express these drag forces as:
\begin{equation}
	\label{eqn:SPH_viscosity_porous}
		\fpexp{\v{f}}{\text{drag}} \left(\v{v}^{n+1} \right) = \tilde{d} \sum_{\insolidneighborhood{\fpID}} \frac{\mu^{\text{por}}}{1 - \phi} \frac{\fpp{m} \sn{m}}{\fpexp{\rho}{n} \snexp{\rho}{0}} \frac{\v{v}_{\fpID \snID}^{n+1} \cdot \v{x}_{\fpID \snID}^{n}}{\|\v{x}_{\fpID \snID}^{n} \|^2 + 0.01 h^2} \nabla  W_{\fpID \snID}^{n},
\end{equation}
where $\mu^{\text{por}}$ is the porous viscosity coefficient and $\tilde{d} = 2(d+2)$ where $d$ is the number of space dimensions.
\eq{eqn:SPH_viscosity_porous} is an approximation of the relative velocity Laplacian, leading to a force similar to the fluid viscosity term in the Navier-Stokes equation (\eq{eqn:navier_stokes}).
The drag forces acting on the solid particles then follow from force balance.
The implicit formulation is inspired by the viscosity method by \citet{Weiler18}, which we modified to achieve symmetric forces between particles of different mass.

\subsubsection{Buoyancy}

Some porous materials float due to the pressure acting on them by the surrounding fluid.
We achieve this effect using pressure forces acting from fluid on solid particles.
Including this force in the implicit pressure solver would allow fluid particles to push away solid particles, without regard for the elastic forces keeping them in their relative configurations.
We therefore consider the solid particles to be fixed when computing the pressure forces for the fluid particles, which results in a more accurate estimate for interactions with relatively stiff solid objects with open pores, where pressure forces are rather resolved by fluid displacement than solid deformation.

Based on the fluid pressure acceleration in \eq{eqn:pressure_accel_fluid}, the sum of the pressure forces acting from solid particle $\spID$ on fluid particles $\fnID$ results in the mirrored force:
\begin{equation}
	\label{eqn:buoyancy_interaction_forces}
	\v{b}_{\spID}^{\text{buo},n} = - (1 - \phi) \spexp{V}{0} \sum_{\influidneighborhood{\spID}} \fn{m} \fnexp{\rho}{0} \left( \frac{\fn{p}}{\fnexp{\hat{\rho}}{2}} \right) \nabla W_{\spID \fnID}.
\end{equation}
Here, we use the pressure forces as determined in the most recent pressure solve step to approximate the fluid pressure gradient, such that this force does not depend on particle velocities.
As this interaction force is already included in the pressure solver for the fluid particles, those do not experience a buoyancy force ($\v{f}_{\fpID}^{\text{buo}} = \v{0}$).

\subsection{Non-Coupling Forces} \label{sec:non-pressure}

For the other terms in the momentum balance equations (\eq{eqn:solid_momentum} and \eq{eqn:navier_stokes}), there already exist many specialized SPH solvers.
In this section we will show how to adapt the implicit fluid viscosity solver by \citet{Weiler18} and the corotated linear elasticity solver by \citet{Peer17} to our porous flow model.
We chose those solvers, since they both use force formulations which are linear in velocity and can therefore be easily combined into a strongly coupled implicit system with our porous interaction forces.

\subsubsection{Fluid Viscosity}
For the fluid viscosity, we use a variation of \eq{eqn:SPH_viscosity_porous} to achieve symmetric forces between the particles:
\begin{equation}
	\label{eqn:SPH_viscosity}
	\fpexp{\v{f}}{\text{vis}} \left(\v{v}^{n+1} \right) = \tilde{d} \sum_{\influidneighborhood{\fpID}} \mu^{\text{vis}} \frac{\fpp{m} \fn{m}}{\fpexp{\hat{\rho}}{n} \fnexp{\hat{\rho}}{n}} \frac{\v{v}_{\fpID \fnID}^{n+1} \cdot \v{x}_{\fpID \fnID}^{n}}{\|\v{x}_{\fpID \fnID}^{n} \|^2 + 0.01 h^2} \nabla  W_{\fpID \fnID}^{n}.
\end{equation}
Here we use the fluid phase densities $\hat{\rho}$ instead of $\rho$, as this is an internal fluid force and we therefore only want to integrate over the domain occupied by the fluid.

\subsubsection{Solid Elasticity}
For the porous elastic solid, we assume that the elasticity coefficients are defined to model the behavior of the whole porous body, including pores.
Then we can apply the elastic solid solver by \citet{Peer17} directly, using $\spexp{V}{0}$ for the particle rest volumes and $\spp{\rho}$ for the density, to model elastic forces:
\begin{equation}
	\label{eqn:SPH_elasticity}
		\frac{1}{\spp{m}}\spexp{\v{f}}{\text{elast}} = \frac{1}{ \spp{\rho}} \nabla \cdot \boldsymbol{\sigma}_{\spID} \left( \v{x}^n, \v{v}^{n+1} \right).
\end{equation}
We refer to the work by \citet{Peer17} for details on how to compute the stress tensor divergence, which can be split into a term that depends linearly on particle velocities $\v{v}^{n+1}$ and a term $\mathbf{b}^{\text{elast},n}$ that only depends on the current particle positions $\v{x}^{n}$.

\subsubsection{Porous Solid Effects}
We further modify the stress tensor computation to include porous material effects like bloating and saturation dependent elastic behavior.
This results in the modified computation of the stress tensor  $\spp{\boldsymbol{\sigma}}$:
\begin{equation}
	\label{eqn:stress}
	\spp{\boldsymbol{\sigma}} = 2 \mu(\spp{S}) \spp{\boldsymbol{\epsilon}} + \lambda(\spp{S}) \text{tr} ( \spp{\boldsymbol{\epsilon}} ) \idMat + \spexp{\boldsymbol{\sigma}}{\text{bloat}}.
\end{equation}

One possible cause for bloating is the increase of volumetric pressure inside filled pores, as the absorbed fluid has a higher pressure than air.
This in turn can cause the pores to expand for soft solids.
For biological materials, like wood or natural sponges, expansion is instead more prevalently caused by hygroscopic swelling \citep{Ha18}.
Here, water not only flows through the pores, but is also absorbed into the cells of the solid skeleton.
The solid itself then increases in volume, observable as a volumetric strain $\epsilon_h = \eta_{h} S$ based on the hygroscopic expansion coefficient $\eta_{h}$.
Following \citet{Lenaerts08}, instead of modifying the strain tensor, we add a bloating term to the stress tensor:
\begin{equation}
	\label{eqn:bloating}
	\spexp{\boldsymbol{\sigma}}{\text{bloat}} = - \eta_{\text{bloat}} \spp{S} \idMat,
\end{equation}
controlled by the bloating factor $\eta_{\text{bloat}}$, which can be understood as a hygroscopic expansion coefficient that is premultiplied with the elasticity parameters.
This term modifies the rest configuration of the solid particles and is therefore independent of the current deformation measured by the strain tensor $\spp{\boldsymbol{\epsilon}}$.
In our implicit solver for the particle velocities, given in Sec.~\ref{sec:system}, it is only included on the right-hand side of the system.

The elastic response of a solid then also depends on the Lamé coefficients $\mu$ and $\lambda$.
Many porous materials change their elastic behavior when absorbing fluid.
One example is wood, which often becomes more malleable with increasing moisture \citep{Persson00}.
For our model, we include this property of porous solids by adjusting the Lamé coefficients based on saturation:
\begin{equation}
	\label{eqn:elasticity_change}
	\mu (\spp{S}) = (1 + \eta_m \spp{S}) \mu^0, \quad \lambda  (\spp{S}) = (1 + \eta_l \spp{S}) \lambda^0,
\end{equation}
with change factors $\eta_m, \eta_l$.
Positive change factors increase and negative ones decrease resistance against deformation, while in the latter case the Lamé coefficients have to be clamped to ensure positivity.
Even though this simplified linear model for elasticity change is motivated purely phenomenologically, it can achieve visually plausible results.
Note that these adjustments can be replaced with more fitting ones for materials where the coefficient change is not monotone or very non-linear.
This is the case for sand, where a very specific amount of added water can greatly increase pile stability, while further increasing saturation causes the grains to flow like a liquid \citep{Pakpour12}.

\subsection{Solver} \label{sec:system}

At the beginning of each time step, we first compute particle densities (\eq{eqn:SPH_density_gen}, \eq{eqn:SPH_density_solid} and \eq{eqn:SPH_density_covered}) and solid particle saturation (\eq{eqn:saturation}).
Then we solve all non-pressure forces to predict the new particle positions.
Solving the porous interaction forces from Section~\ref{sec:porous-forces} in an explicit fashion becomes highly unstable for large drag forces.
We therefore propose to use an implicit solver.

Given the mass matrix $\v{M}$ and forces $\v{f}$ which are linear in $\v{v}^{n+1}$, we write the velocity update as:
\begin{equation}
	\label{eqn:velocity_update}
	\begin{aligned}
	%\v{M} \v{v}^{n+1} &\approx \v{M} \v{v}^n + \Delta t \v{f}\left( \v{x}^n, \v{v}^{n+1} \right) \\
	\v{M} \v{v}^{n+1} - \Delta t \v{f}\left( \v{v}^{n+1}\right) \approx \v{M} \v{v}^{n}  + \Delta t \v{b}^{n}.
	\end{aligned}
\end{equation}
At timestep $n$ we solve this linear system for $\v{v}^{n+1}$, which is the stacked vector of all particle velocities for the next time step.
The total force $\v{f}$ and right-hand side $\v{b}$ are the sums of all the corresponding terms in Section~\ref{sec:porous-forces} and Section~\ref{sec:non-pressure}.
Due to our choice of discretization, these forces can readily be combined into one linear system and solved using a matrix-free conjugate gradient method.
More details about this linear system are given in our supplementary document.

As proposed by \citet{Weiler18}, we add the velocity updates $\v{v}^{n} - \v{v}^{n-1}$ from the last timestep to the current velocities $\v{v}^{n}$ to serve as an initial guess when solving the system for $\v{v}^{n+1}$.
After the new velocities are obtained, particle densities are predicted using \eq{eqn:SPH_density_change}, which are corrected by pressure accelerations, as described in Section~\ref{sec:pressure-solver}.
These accelerations in conjunction with the velocities $\v{v}^{n+1}$ are then integrated using the symplectic Euler method, resulting in new particle positions for the next time step.
\section{Results}

\begin{figure}
	\centering
	\begin{subfigure}{\linewidth}
		\caption{Explicit}
		\begin{subfigure}{.5\linewidth}
			\centering
			\includegraphics[width=0.95\linewidth,trim={800 140 700 600},clip]{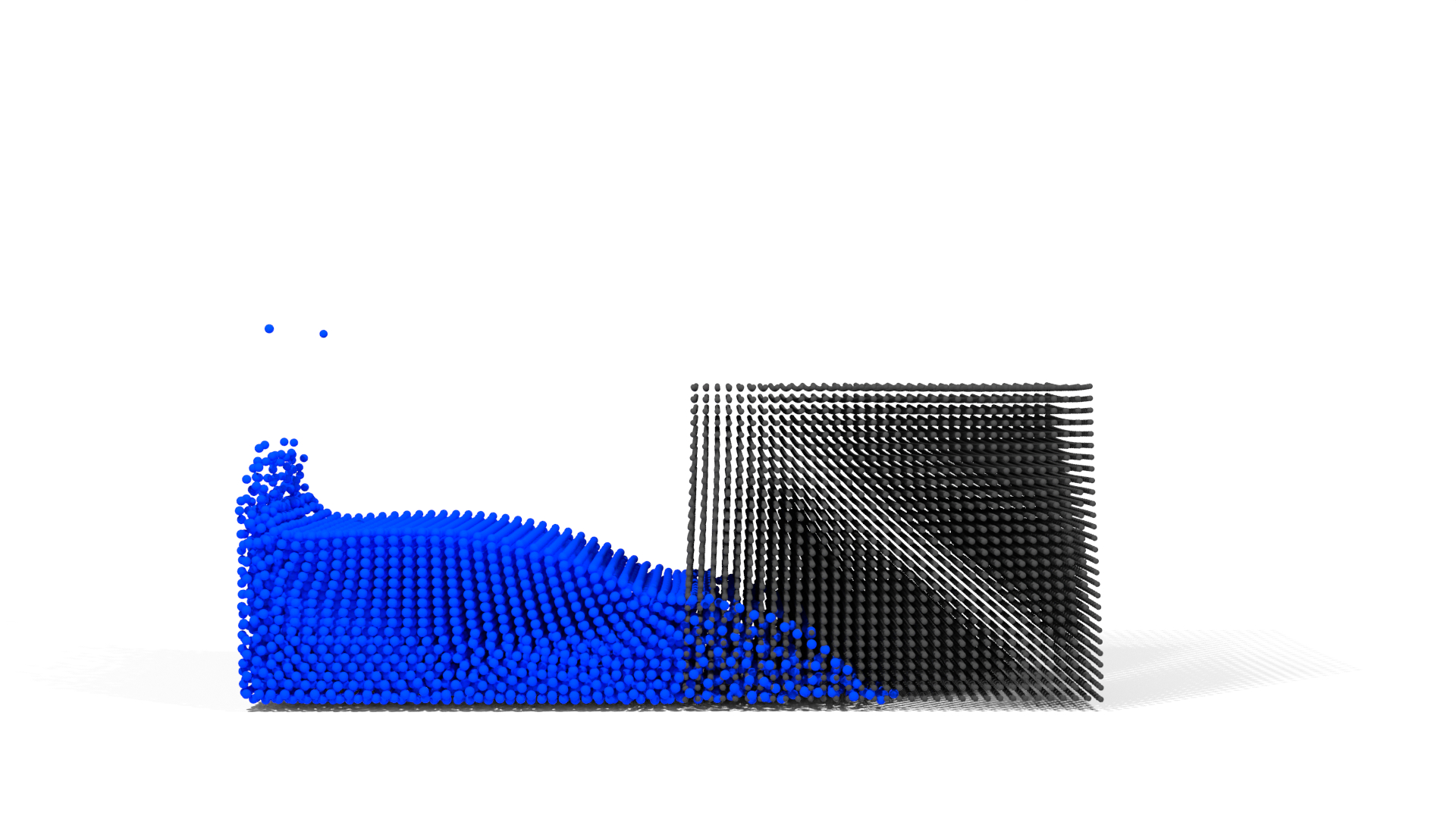}
		\end{subfigure}%
		\begin{subfigure}{.5\linewidth}
			\centering
			\includegraphics[width=0.95\linewidth,trim={800 140 700 600},clip]{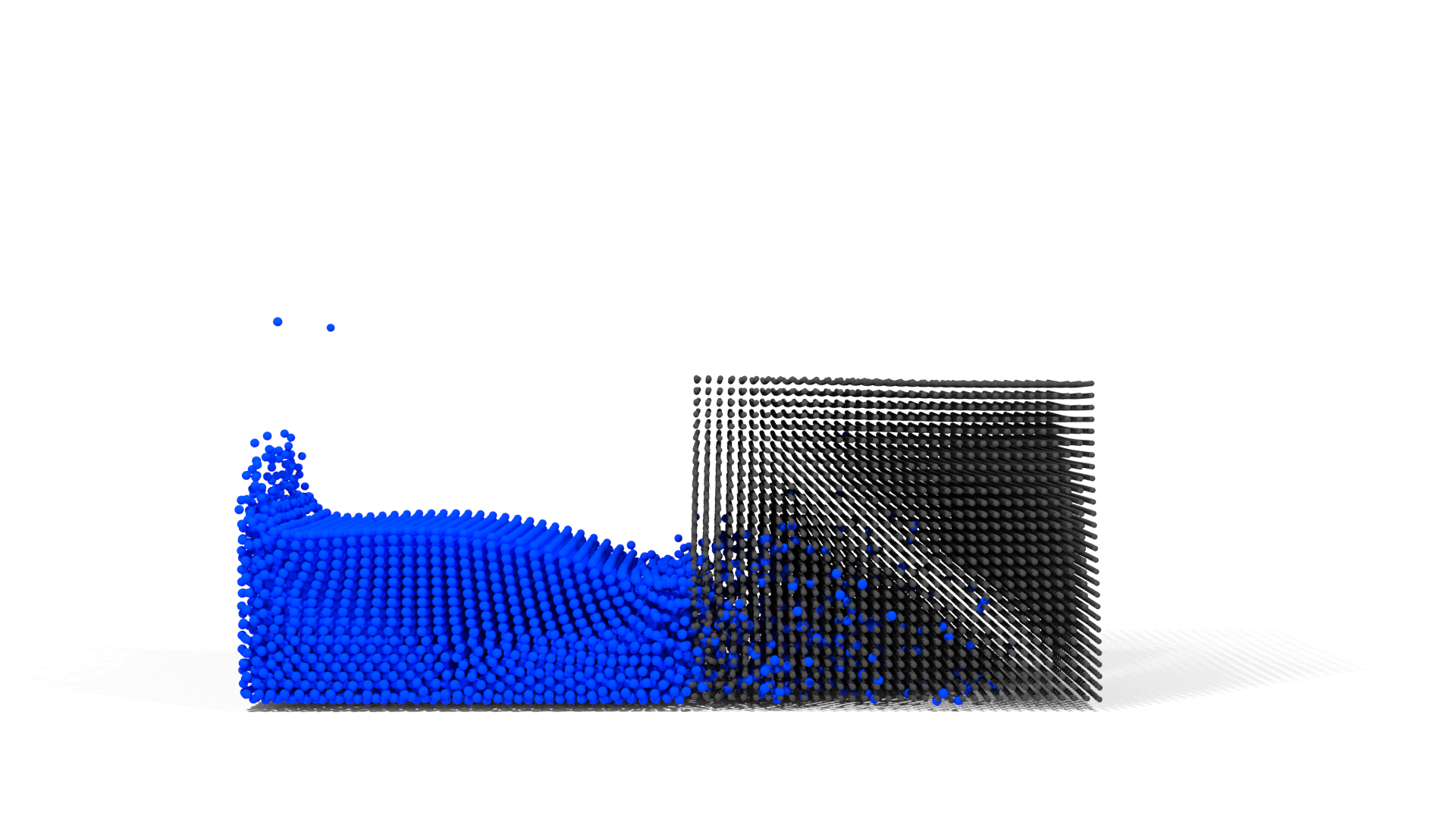}
		\end{subfigure}
	\end{subfigure}
	\begin{subfigure}{\linewidth}
		\caption{Implicit}
		\begin{subfigure}{.5\linewidth}
			\centering
			\includegraphics[width=0.95\linewidth,trim={800 140 700 600},clip]{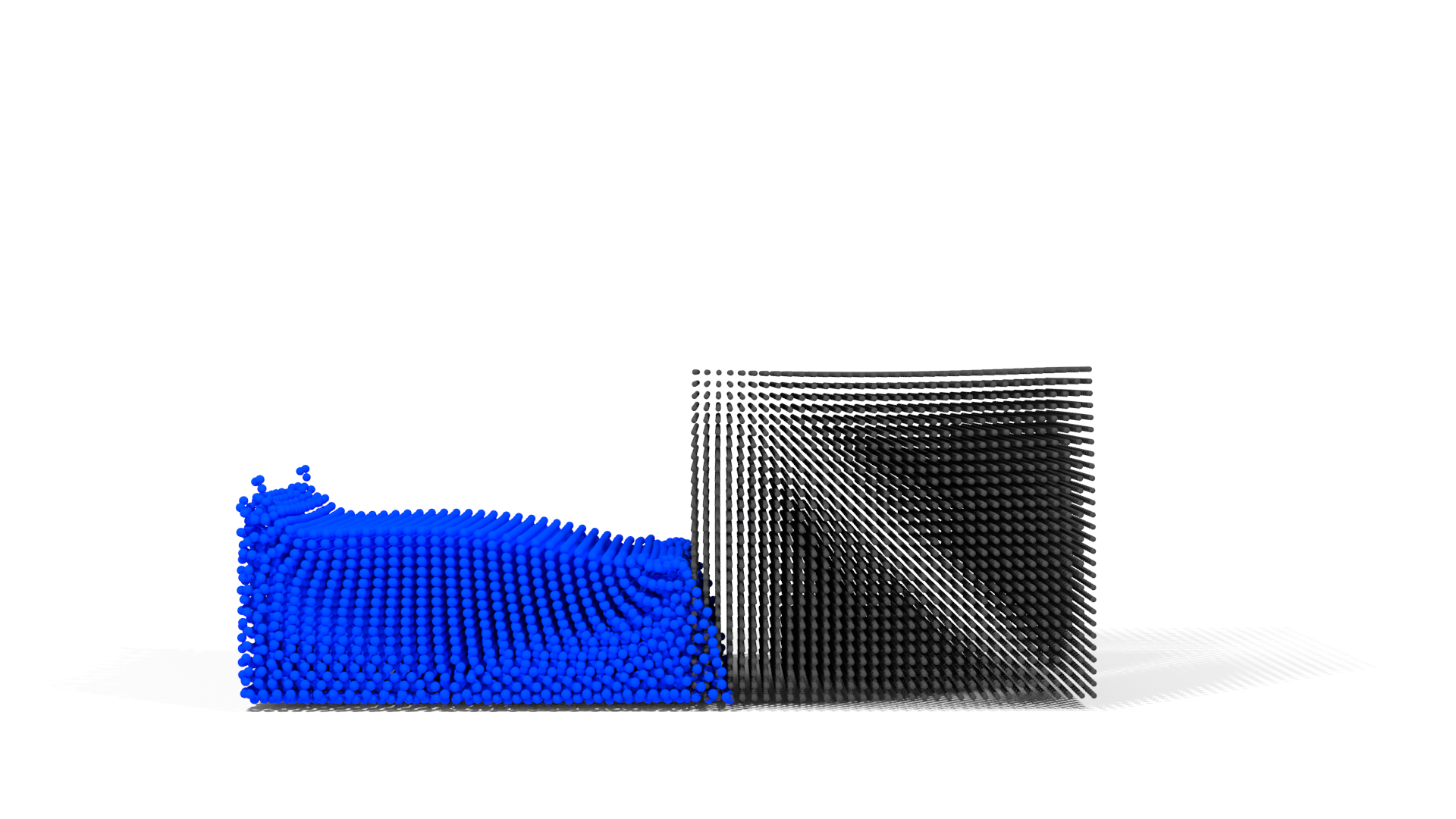}
		\end{subfigure}%
		\begin{subfigure}{.5\linewidth}
			\centering
			\includegraphics[width=0.95\linewidth,trim={800 140 700 600},clip]{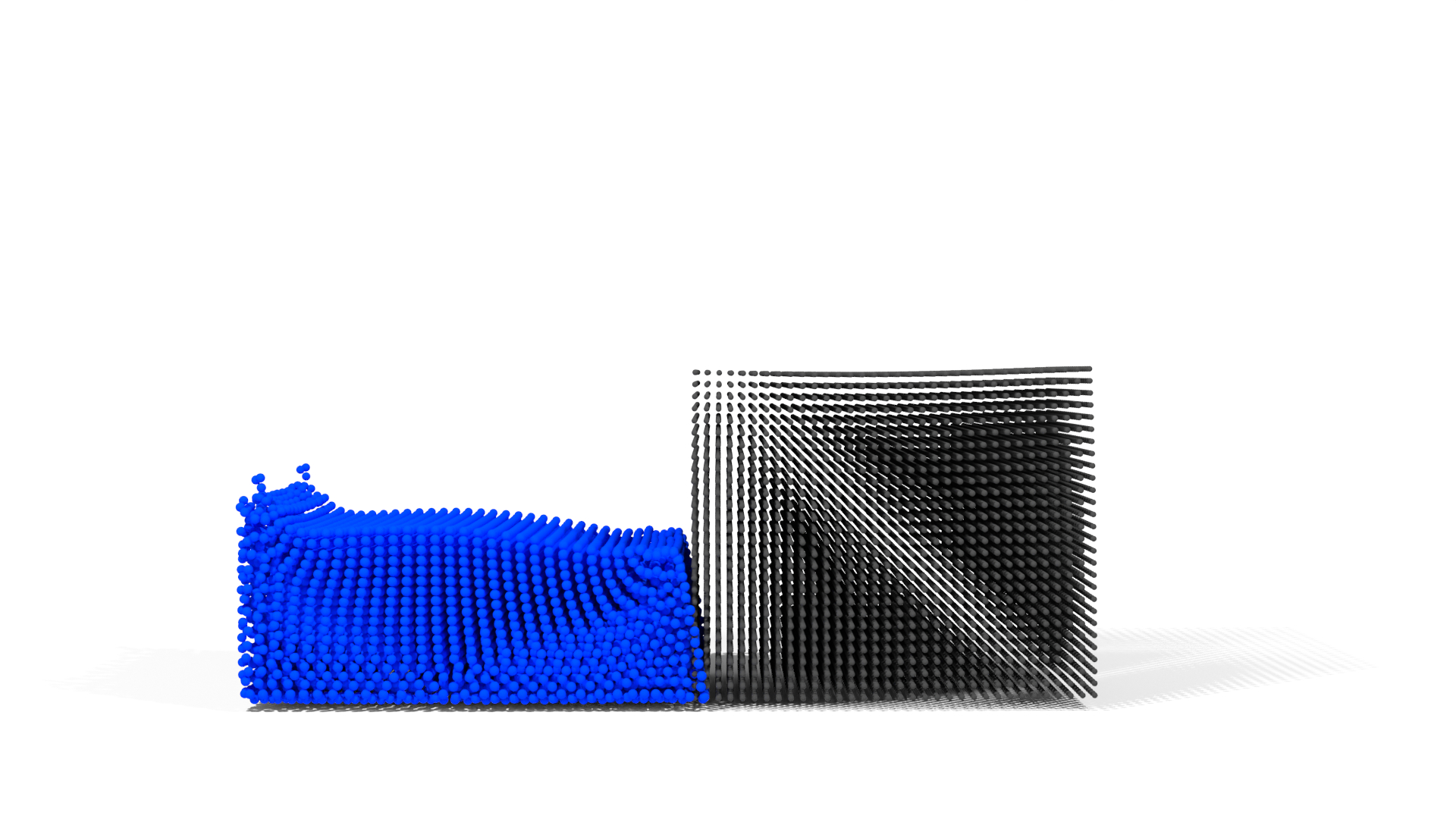}
		\end{subfigure}
	\end{subfigure}
	\caption{Fluid (blue) interacts with a porous wall (gray, particles scaled down for  internal visibility). Shown is the state $0.12 \si{\second}$ after the water first comes into contact with the solid, using explicit or implicit coupling forces. The left two examples use a porous viscosity coefficient of $\mu^{\text{por}} = 10 \si{\pascal \second}$ while the right two use $\mu^{\text{por}} = 100 \si{\pascal \second}$. For a constant time step size of $1 \si{\milli \second}$ explicit forces cannot sufficiently slow down fluid flow with the smaller coefficient and are unstable for larger one.}
	\label{fig:wall}
\end{figure}

\begin{figure}
	\begin{subfigure}{\linewidth}
		\caption{\citet{Ren21}}
		\begin{subfigure}{.48\linewidth}
			\includegraphics[width=\linewidth,trim={750 140 500 300},clip]{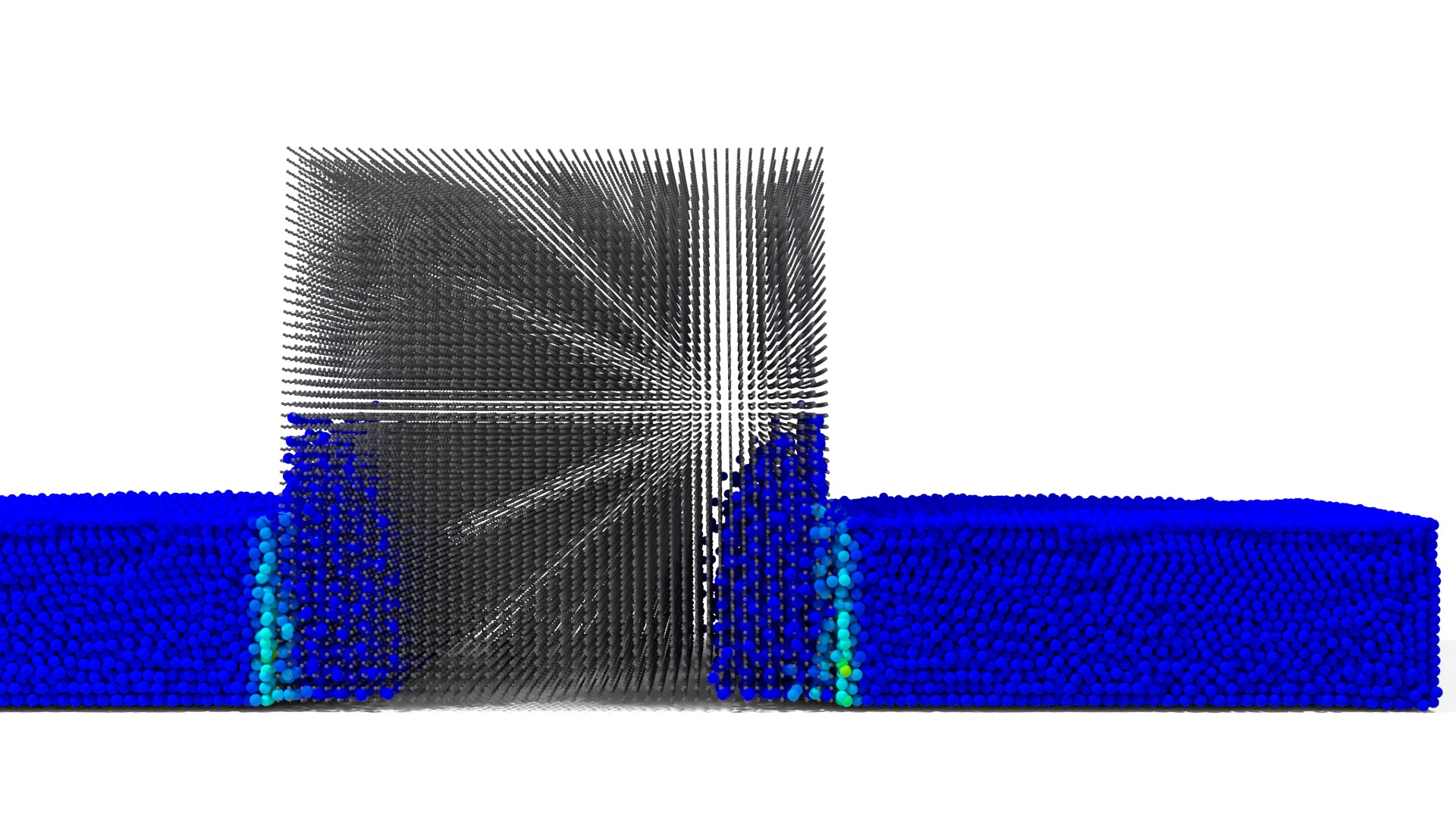}
		\end{subfigure}
		\begin{subfigure}{.48\linewidth}
			\includegraphics[width=\linewidth,trim={750 140 500 300},clip]{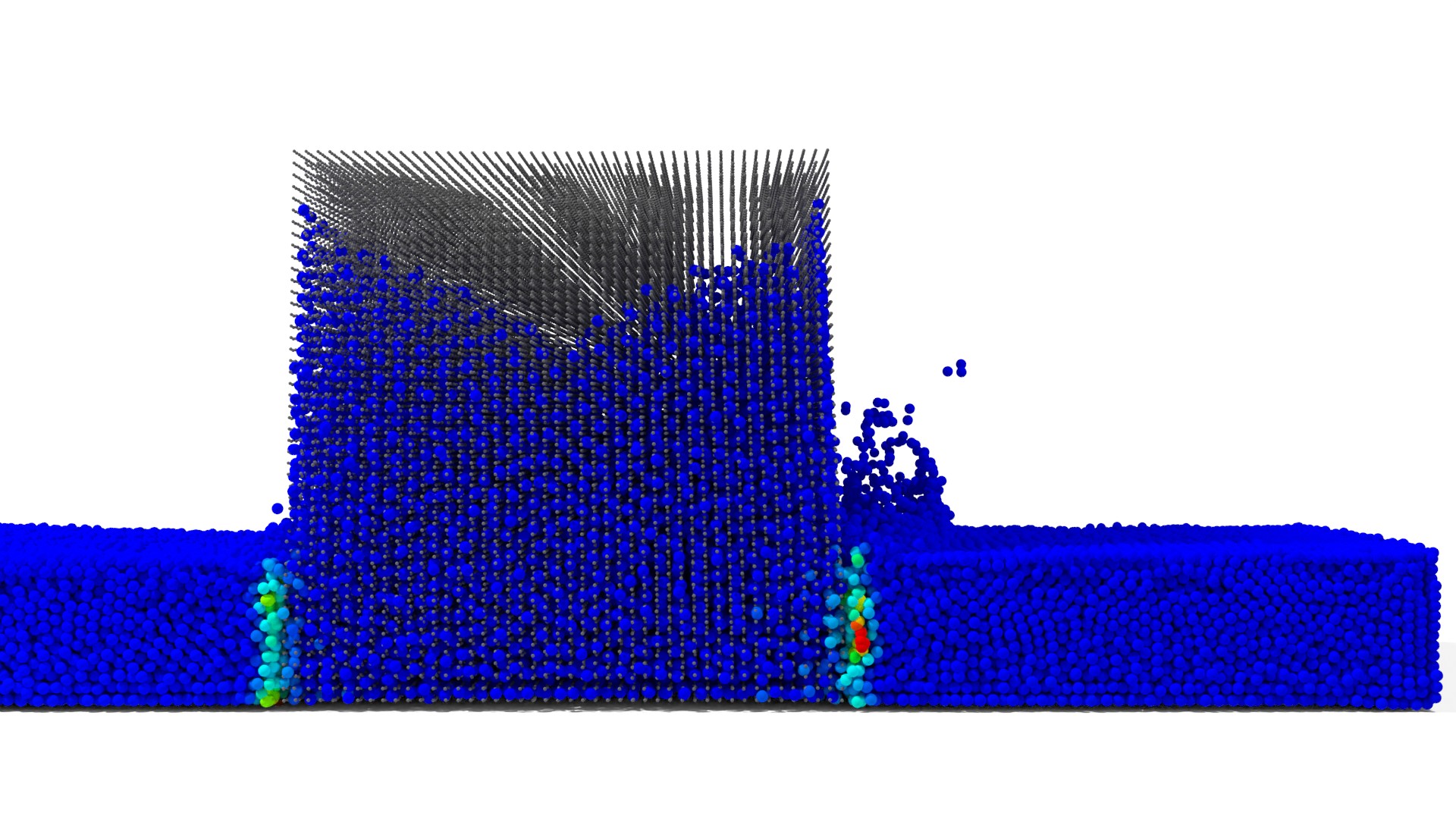}
		\end{subfigure}
	\end{subfigure} \\
	\begin{subfigure}{\linewidth}
		\caption{Ours}
		\begin{subfigure}{.48\linewidth}
			\includegraphics[width=\linewidth,trim={750 140 500 300},clip]{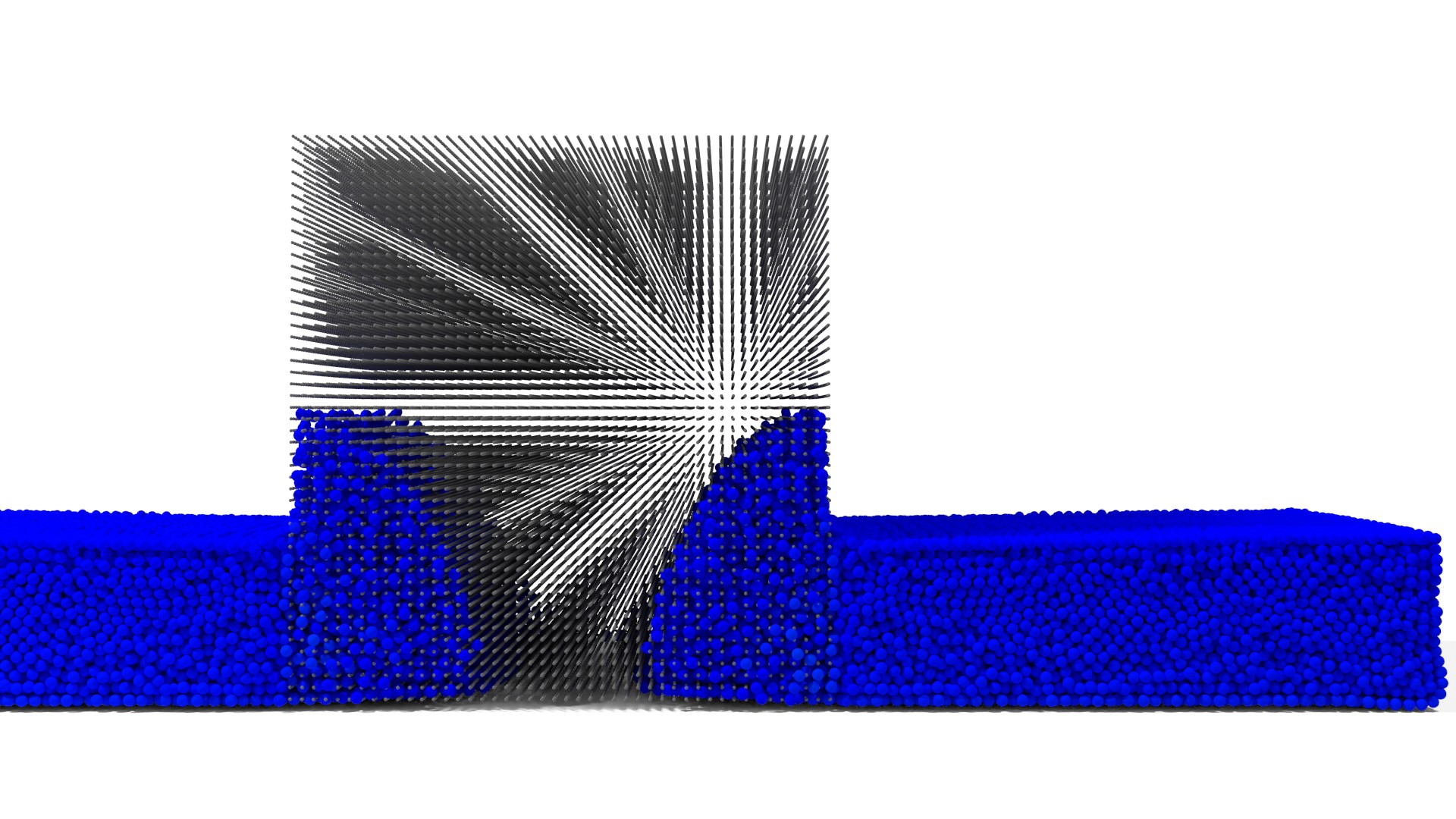}
		\end{subfigure}
		\begin{subfigure}{.48\linewidth}
			\includegraphics[width=\linewidth,trim={750 140 500 300},clip]{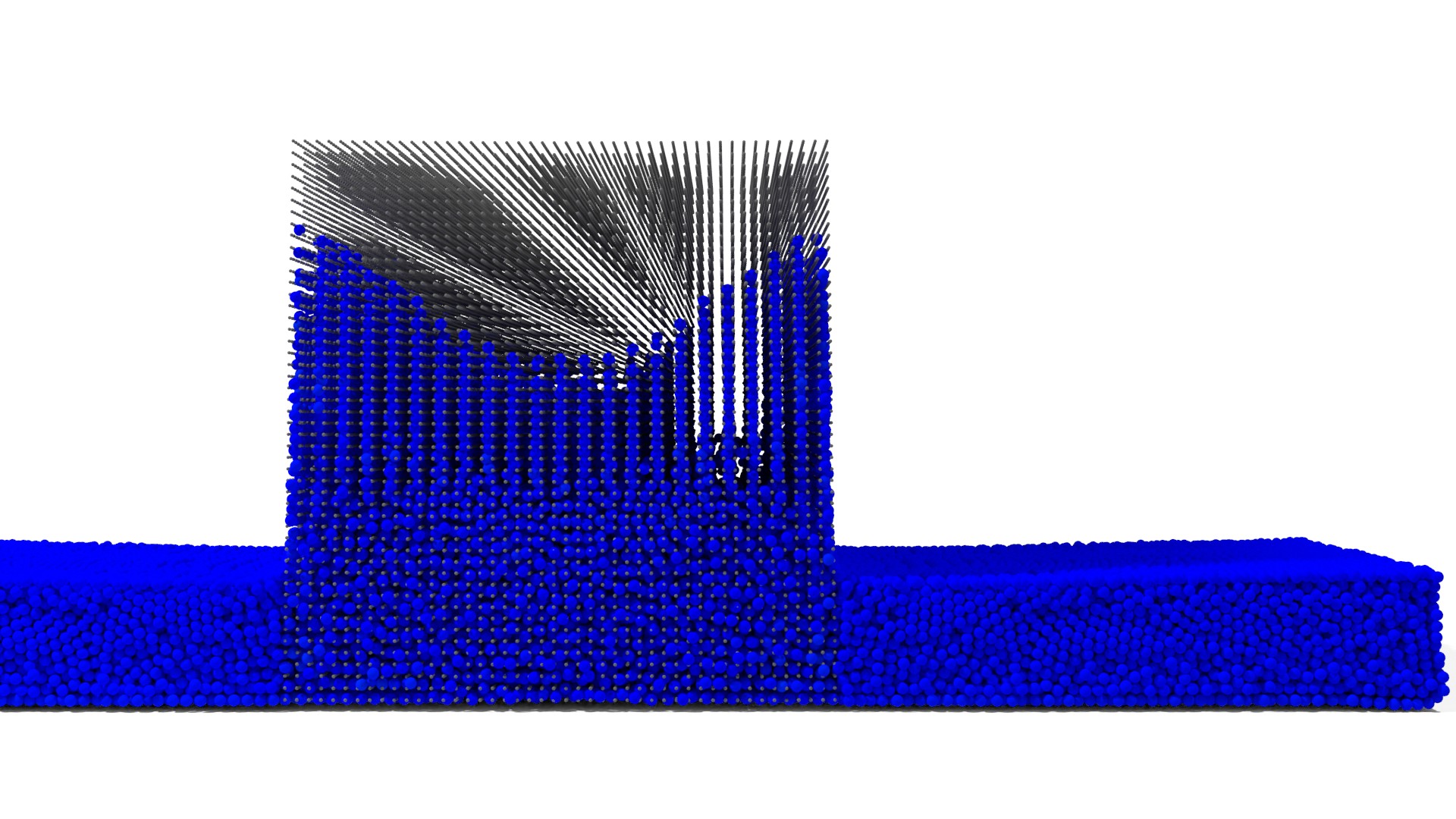}
		\end{subfigure}
	\end{subfigure} 
	\caption{Cut plane view of water being absorbed into a porous object (gray, particles scaled down for  internal visibility) with significant capillary action. The fluid is colored according to the phase density $\hat{\rho}$, ranging from $\leq 1000$ (blue) to $\geq 2000$ \si{\kilogram\per\cubic\meter} (red). Shown is the state after $3 \si{\second}$ for slow (left) and fast (right) porous flow. For the fast version, the method by \citet{Ren21} exhibits frequent explosions originating from densely packed particle clusters. Our method can stably simulate either case.}
	\label{fig:adhesion_ren}
\end{figure}

The following results were simulated on an AMD Ryzen Threadripper PRO 5975WX CPU with 32 cores, 3.60 GHz base core frequency and 256 GB of RAM.
We have implemented our method into SPlisHSPlasH~\cite{SPlisHSPlasH_Library}, an SPH fluid simulation library written in C++, and plan to release our code as open source.
To increase performance, we employ AVX vectorization and parallelization using OpenMP.
Surface reconstructions were done using the splashsurf library \citep{LBJB23} or mesh skinning as described by \citet{KBF+21}.
We assume that the porous object is incompressible if not stated otherwise.
Additional simulation parameter choices as well as runtime measurements are given in the supplementary document.

\subsection{Stability}

Using the explicit formulation of \eq{eqn:SPH_viscosity_porous} (i.e., replacing $\v{v}^{n+1}$ with $\v{v}^{n}$) we found that for a given time step size of $1 \si{\milli \second}$ and $\mu^{\text{por}} = 10 \si{\pascal \second}$, the resulting force cannot sufficiently reduce the fluid particle speed upon contact with the porous solid and they instead immediately permeate far inside the block, as shown in \fig{fig:wall}.
Increasing the viscosity coefficient does not solve this issue and instead leads to instabilities.
Implicit forces on the other hand can notably slow down the fluid particles even with the lower coefficient and are still stable for the higher one.

\begin{figure*}
	\begin{subfigure}{.33\linewidth}
		\includegraphics[width=0.99\linewidth,trim={600 160 600 500},clip]{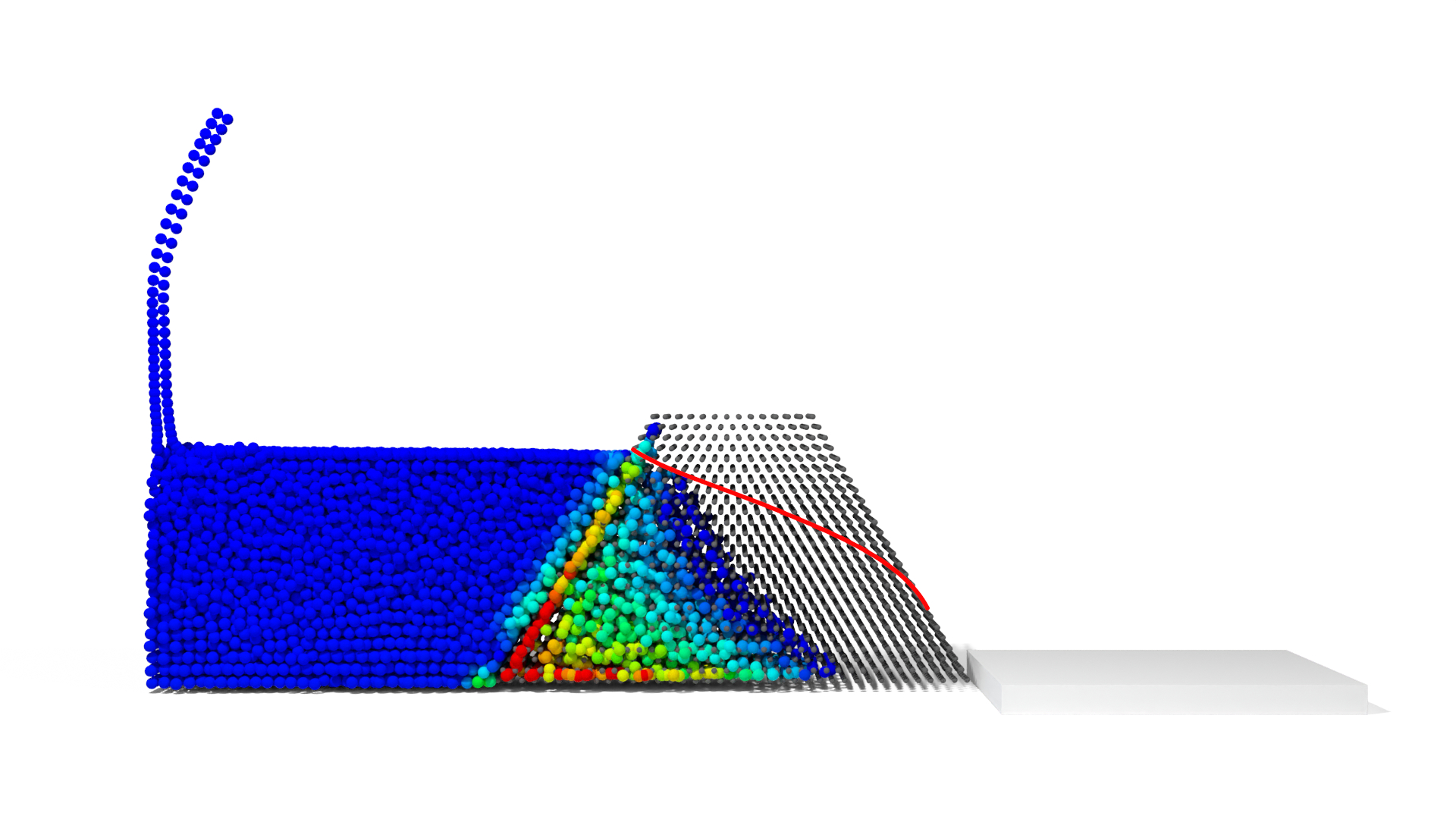}
		\caption{\citet{Ren21}}
	\end{subfigure} 
	\begin{subfigure}{.33\linewidth}
		\includegraphics[width=0.99\linewidth,trim={600 160 600 500},clip]{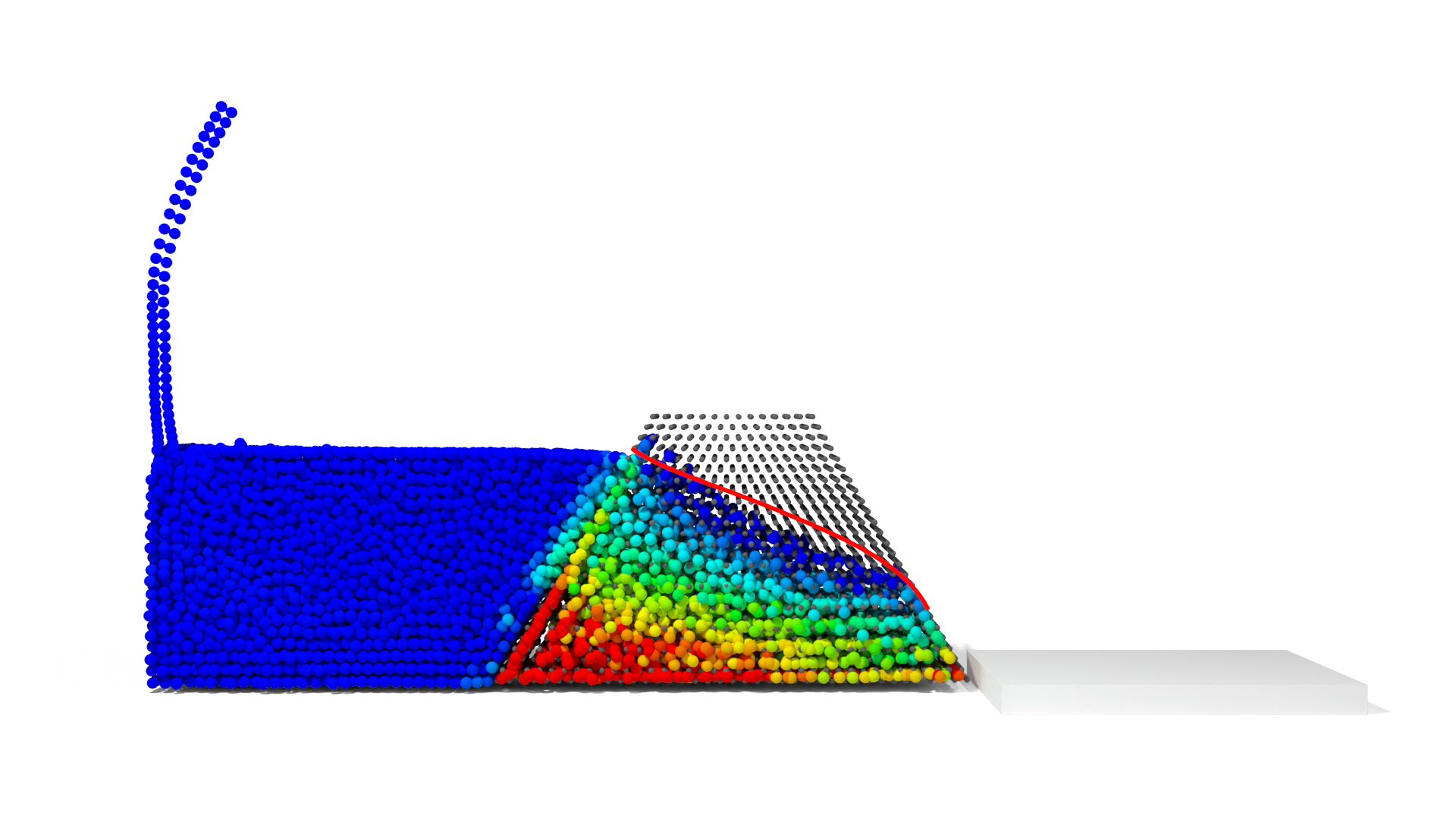}
		\caption{\citet{Ren21} (modified)}
	\end{subfigure} 
	\begin{subfigure}{.33\linewidth}
		\includegraphics[width=0.99\linewidth,trim={600 160 600 500},clip]{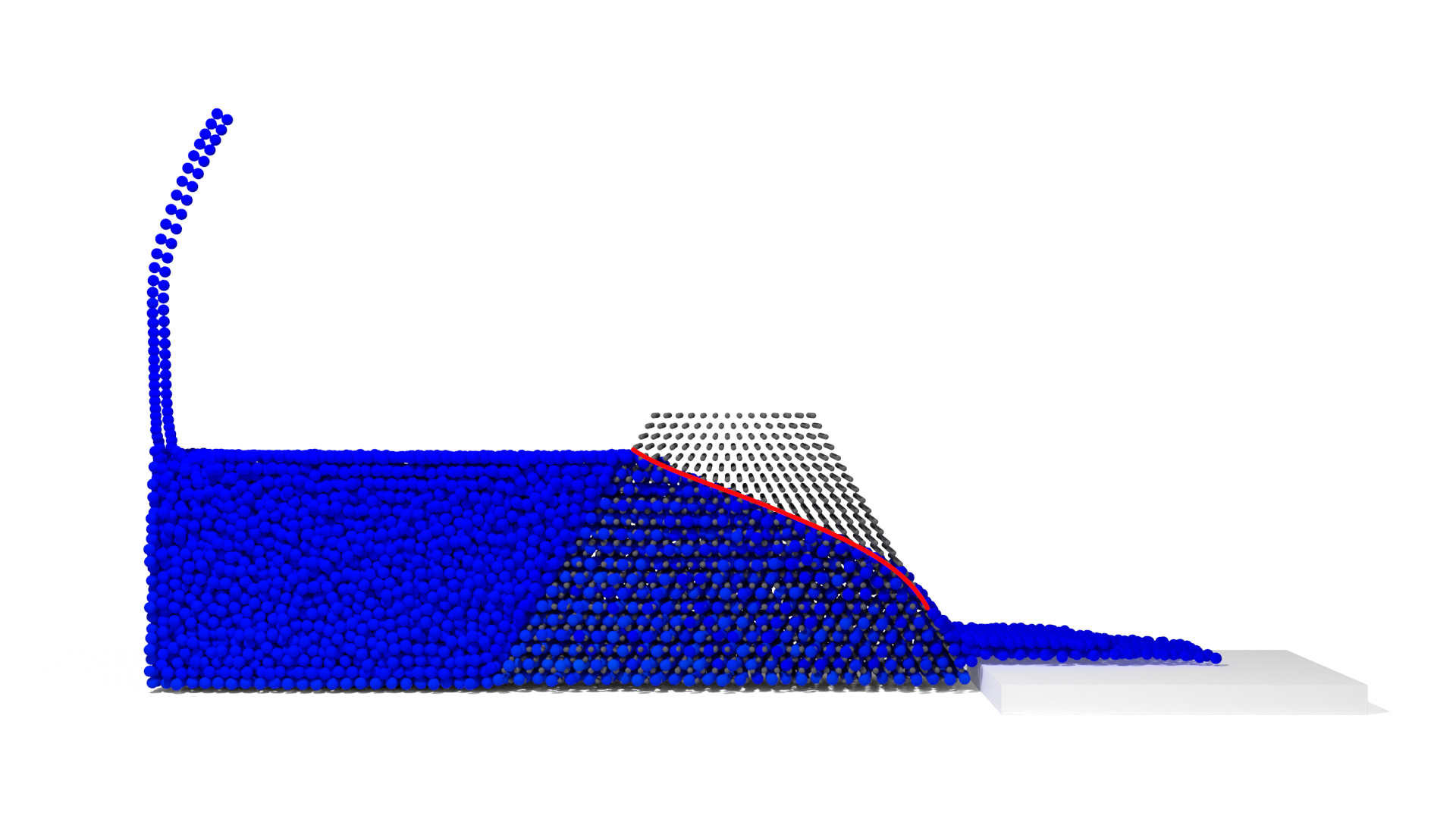}
		\caption{Ours}
	\end{subfigure}
	\caption{Cut-plane view of fluid flowing through a porous dam (gray, particles scaled down for internal visibility), from left to right: Method by \citet{Ren21} with low pore pressure ($p_0 = 500 \si{\pascal}$), with modified Darcy's law ($p_0 = 2000 \si{\pascal}$) and our method without capillary action ($C^{\text{cap},0} = 0 \si{\newton \per \meter}$). Fluid particles are colored according to fluid phase density, ranging from blue ($\fpp{\hat{\rho}} \leq$ 1000 \si{\kilogram\per\cubic\meter}) to red ($\fpp{\hat{\rho}} \geq$ 2000 \si{\kilogram\per\cubic\meter}). The red line shows the Casagrande solution for the expected fluid surface.}
	\label{fig:dam}
\end{figure*}

In \fig{fig:adhesion_ren} we show simulations of a cube ($\phi = 0.6$) placed between two shallow pools of water, where only one half is shown.
We chose a significant capillary action ($C^{\text{cap},0} = 1000 \si{\newton \per \meter}$) to compare our work with the model by \citet{Ren21} ($p_0 = 10000 \si{\pascal}$).
While our method can stably simulate both slow ($k = 1e^{-6}$ or $\mu^{\text{por}} = 50 \si{\pascal \second}$) and fast ($k = 1e^{-5}$ or $\mu^{\text{por}} = 3 \si{\pascal \second}$) porous flow, the method by \citet{Ren21} suffers from pressure spikes originating in dense fluid particle clumps at the solid-fluid interface, where the particles are scaled according to the absorption model. 
This results in explosions when particles rapidly increase in size, which happens more frequently in scenarios with fast porous flow.
While this issue could be addressed with stabilization techniques, e.g., by introducing an upper limit on how much particles are allowed to grow in a time step, our method does not have these concerns since we completely avoid severe fluid particle scaling.

On the other hand, their method is faster, which can be attributed to the missing pressure computations for absorbed fluid particles and the only one-way coupled explicit solver for the porous flow.
For the example with slow flow, our method took on average $194 \si{\milli \second}$ per time step, while the the method by \citet{Ren21} only took $126 \si{\milli \second}$.
Note that, while our reimplementation of their method closely follows code published by the authors, we simulated these examples on the CPU without AVX vectorization, such that their method could potentially perform even faster.

\subsection{Validation}

We further investigated the well-researched use case of fluid transport through a trapezoidal dam, shown in \fig{fig:dam}.
This scenario does not exhibit obvious capillary action, which is difficult to simulate using the method by \citet{Ren21}, since we could not control porous flow speed independently of capillary height due to their definition of the pore pressure.
We therefore extend their approach with a version of Darcy's law that includes gravitational forces: $\v v_{fs} = - \frac{k}{\phi \mu}(\nabla p - \rho_f \v g)$, for which the derivation is given in the supplementary document.
While this allows control over the upwards pull on the absorbed fluid through the rest pore pressure $p_0$, we still could not find parameters that produce the expected result, which we determine using the Casagrande solution \citep{Casagrande1937} for the fluid surface.
Additionally, the fluid does not leave the dam at the opposite side and is instead greatly compressed at the bottom, due to the lack of proper pressure forces acting on absorbed fluid particles.
Our method on the other hand can easily handle the common case of porous flow without capillary action and produces the expected result.

\begin{figure}
	\begin{subfigure}{.32\linewidth}
		\includegraphics[width=\linewidth,trim={700 120 750 400},clip]{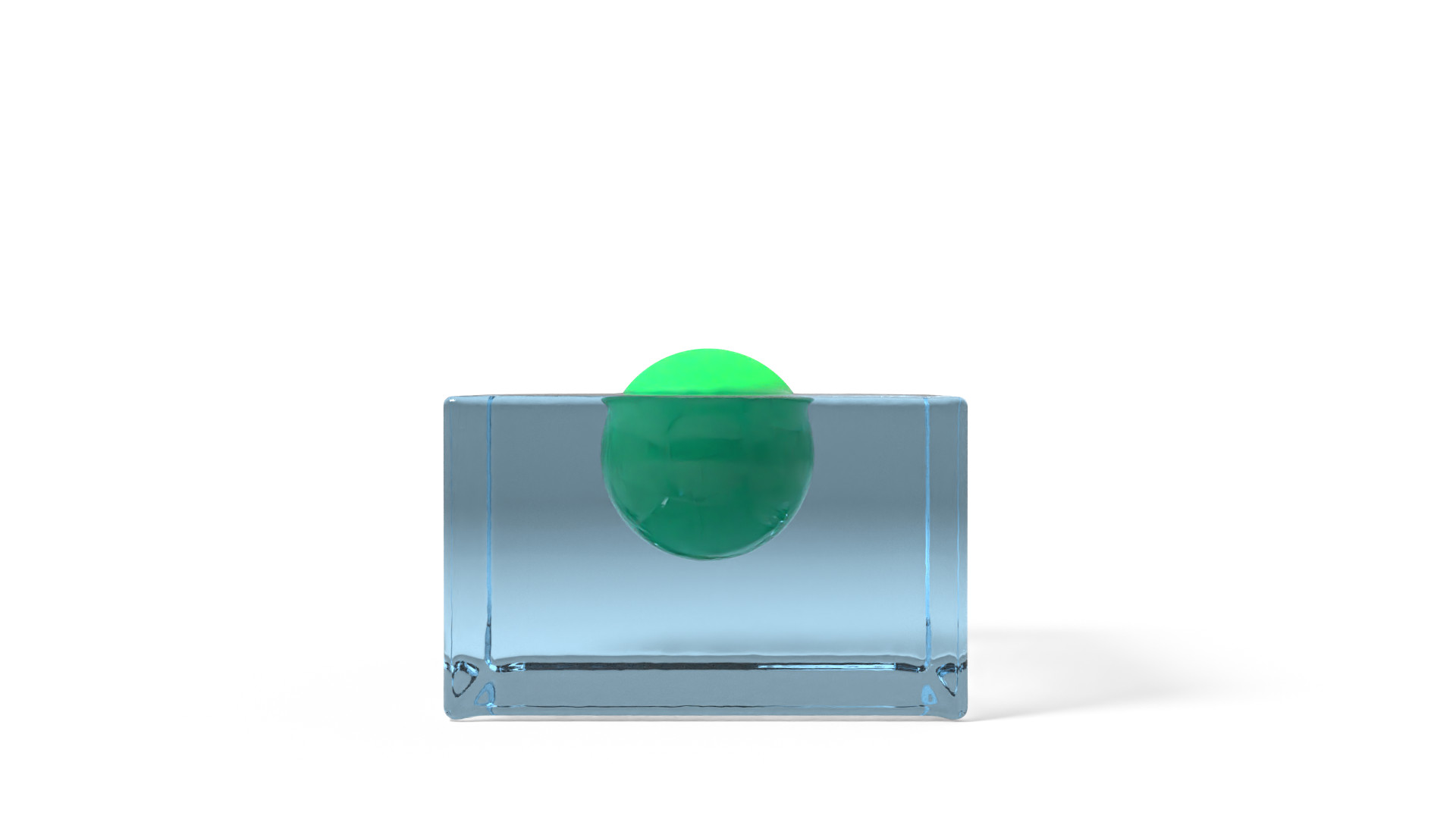}
	\end{subfigure}
	\begin{subfigure}{.32\linewidth}
		\includegraphics[width=\linewidth,trim={700 120 750 400},clip]{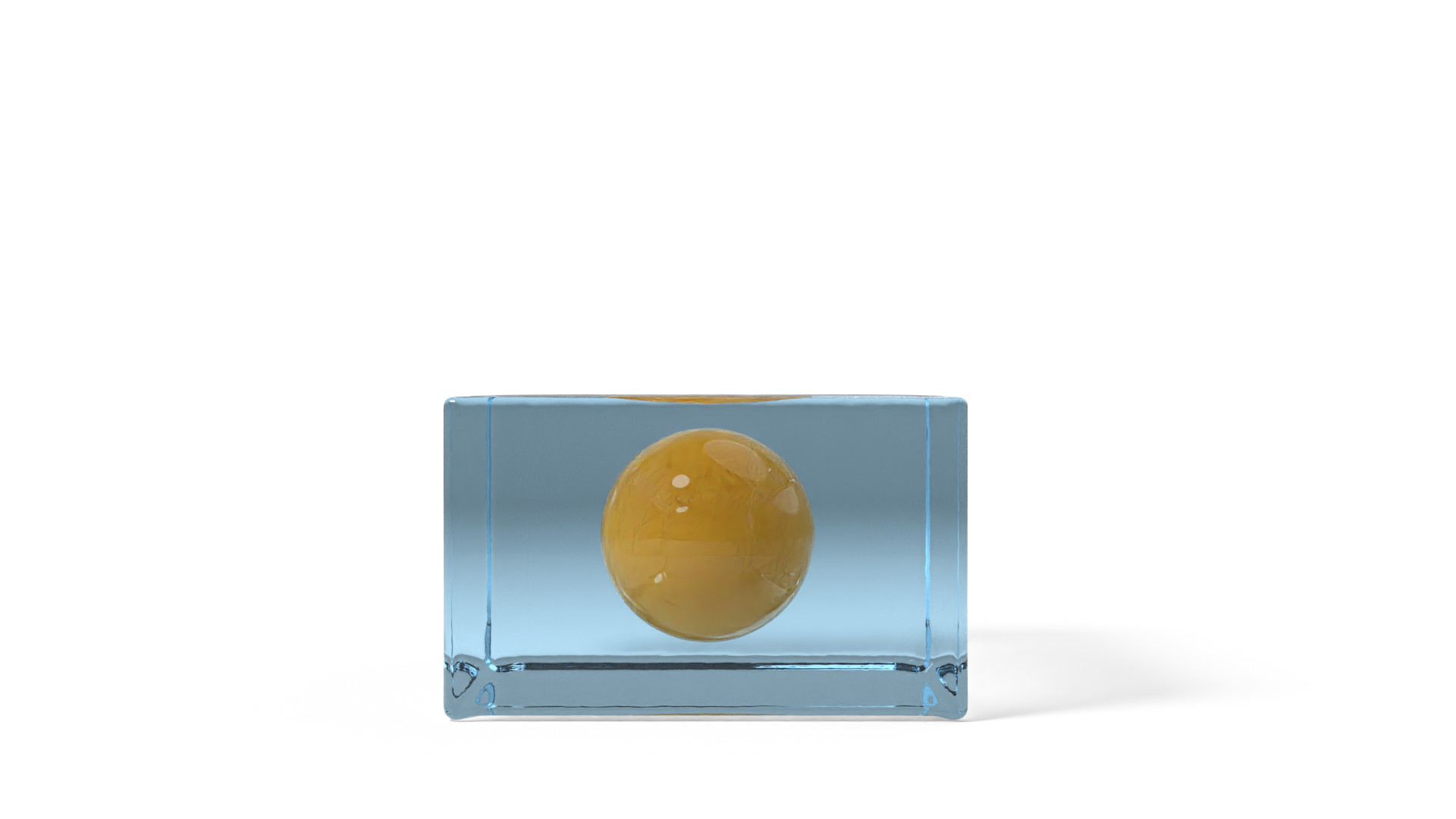}
	\end{subfigure}
	\begin{subfigure}{.32\linewidth}
		\includegraphics[width=\linewidth,trim={700 120 750 400},clip]{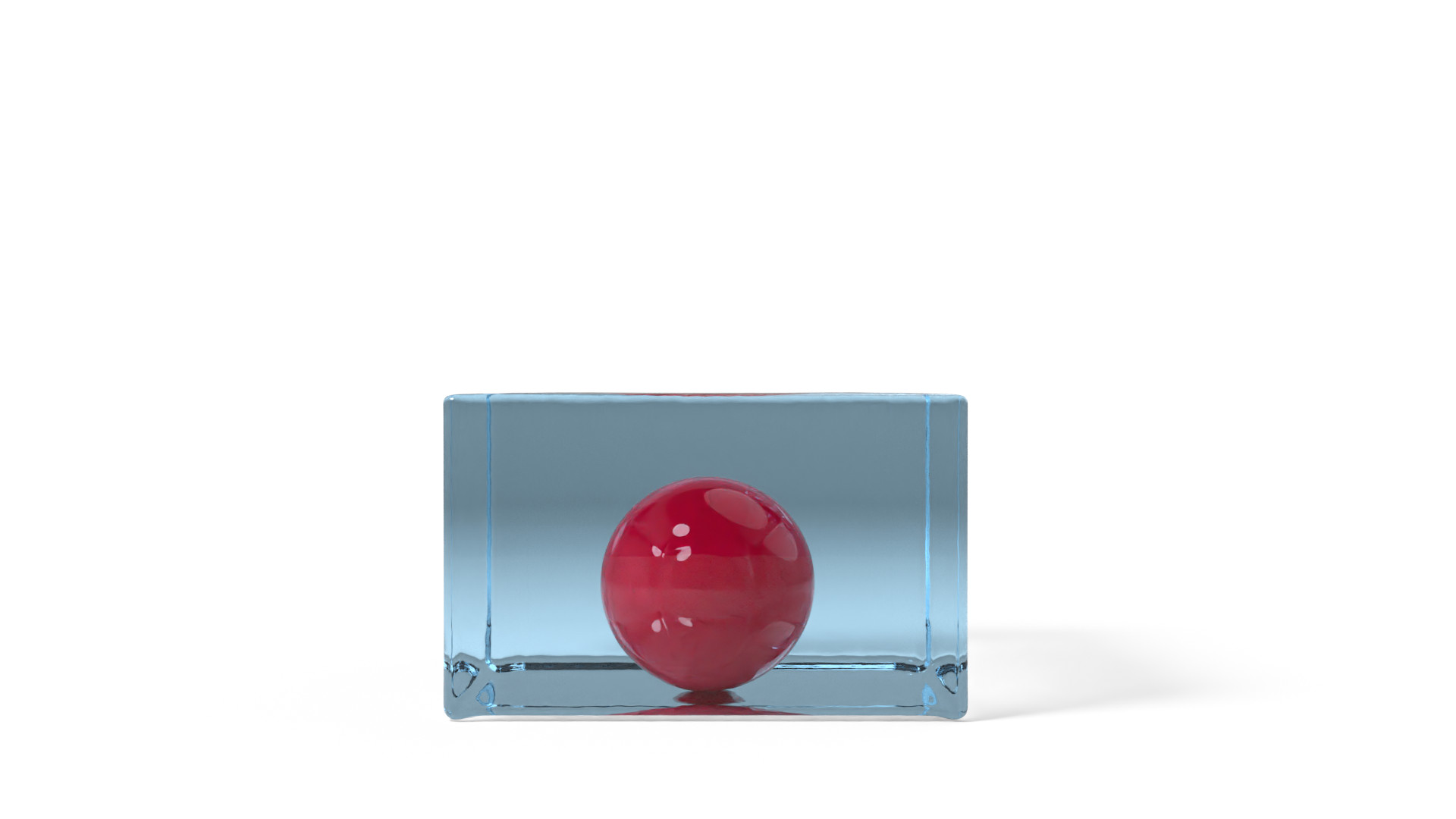}
	\end{subfigure}
	\caption{Porous spheres being dropped in water after $8 \si{\second}$, with varying densities $\rho_{\text{Sphere}} = (1 - \phi) \rho_{\text{Solid}}$. The solid phase density increases from left to right: $\rho_{\text{Solid}} = 888.9 \si{\kilogram \per \cubic \meter}$ (with $\rho_{\text{Sphere}} = 400 \si{\kilogram \per \cubic \meter}$ and $\phi = 0.55$), $\rho_{\text{Solid}} = 1142.9 \si{\kilogram \per \cubic \meter}$ (with $\rho_{\text{Sphere}} = 400 \si{\kilogram \per \cubic \meter}$ and $\phi = 0.65$), and $\rho_{\text{Solid}} = 1555.6 \si{\kilogram \per \cubic \meter}$ (with $\rho_{\text{Sphere}} = 700 \si{\kilogram \per \cubic \meter}$ and $\phi = 0.55$). Only the first sphere floats due to the solid being less dense than water, while the middle sphere sinks slower than the right one.}
	\label{fig:buoyancy}
\end{figure}

Furthermore, we showcase the simulation of objects sinking due to absorbed fluid mass.
\fig{fig:buoyancy} shows porous spheres being dropped in a water bath, where our method correctly captures their expected floating behavior based on the solid phase density.
Only the first sphere floats, while the third sphere sinks faster than the second one due to the smaller buoyancy force.

\subsection{Parameter Studies}

\begin{figure}
	\begin{subfigure}{.24\linewidth}
		\includegraphics[width=\linewidth,trim={90 0 20 0},clip]{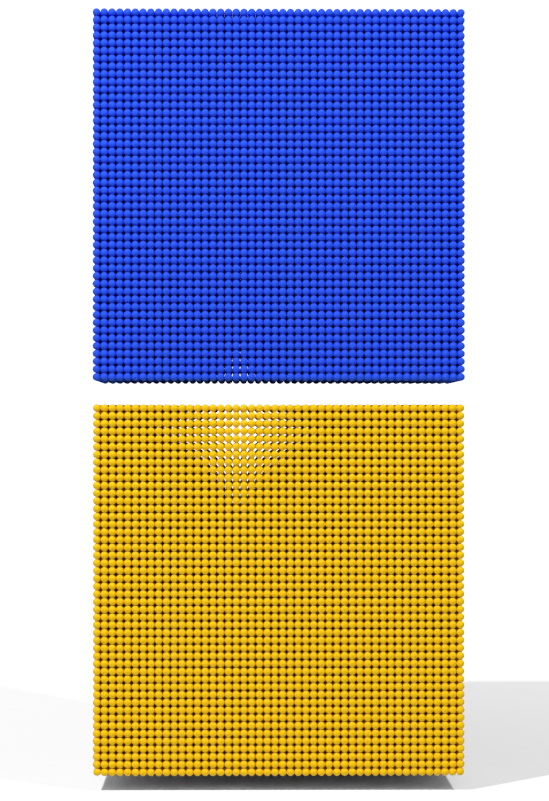}
		\caption{Initial $\quad$}
	\end{subfigure}
	\begin{subfigure}{.24\linewidth}
		\includegraphics[width=\linewidth,trim={90 0 20 0},clip]{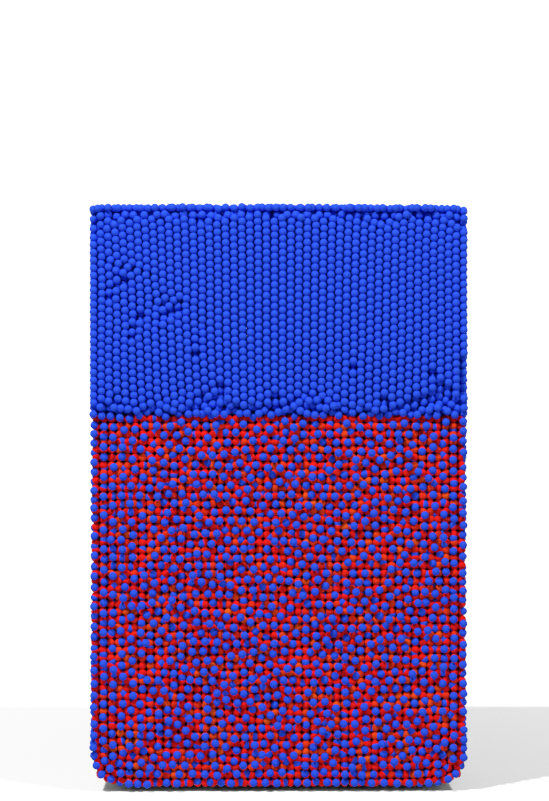}
		\caption{$\phi = 0.4 \quad$}
	\end{subfigure}
	\begin{subfigure}{.24\linewidth}
		\includegraphics[width=\linewidth,trim={90 0 20 0},clip]{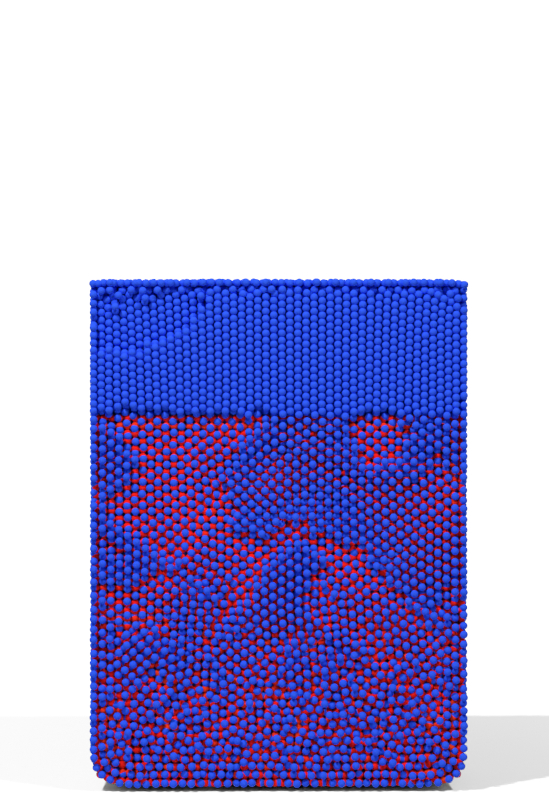}
		\caption{$\phi = 0.6 \quad$}
	\end{subfigure}
	\begin{subfigure}{.24\linewidth}
		\includegraphics[width=\linewidth,trim={90 0 20 0},clip]{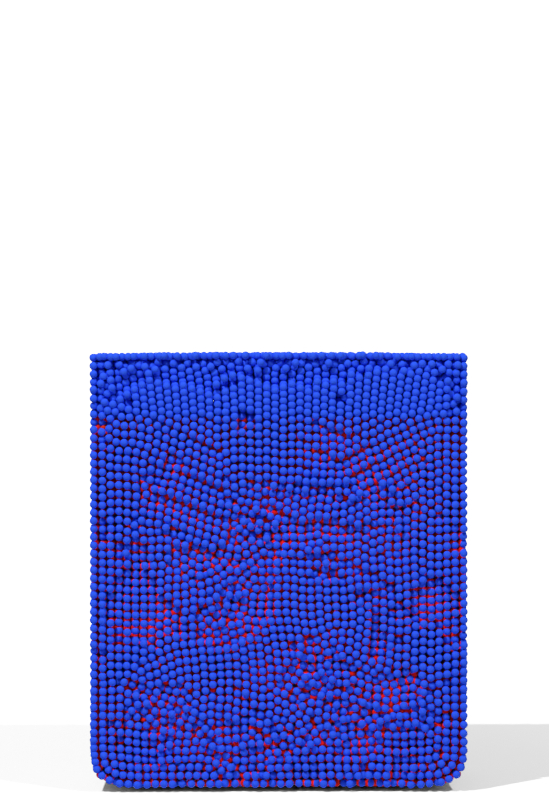}
		\caption{$\phi = 0.8 \quad$}
	\end{subfigure}
	\caption{Internal view of fluid (blue) seeping into porous material, where the particle colors range from no saturation (yellow) to fully saturated (red). Shown is the initial state, followed by the state after $10 \si{\second}$ for $\phi = 0.4, 0.6$, and $0.8$, respectively, where the porous solid is fully saturated. The amount of fluid that can be absorbed directly corresponds to the solid porosity.}
	\label{fig:seep_porosity}
\end{figure}

\begin{figure}
\includegraphics[width=\linewidth,trim={0 130 0 230},clip]{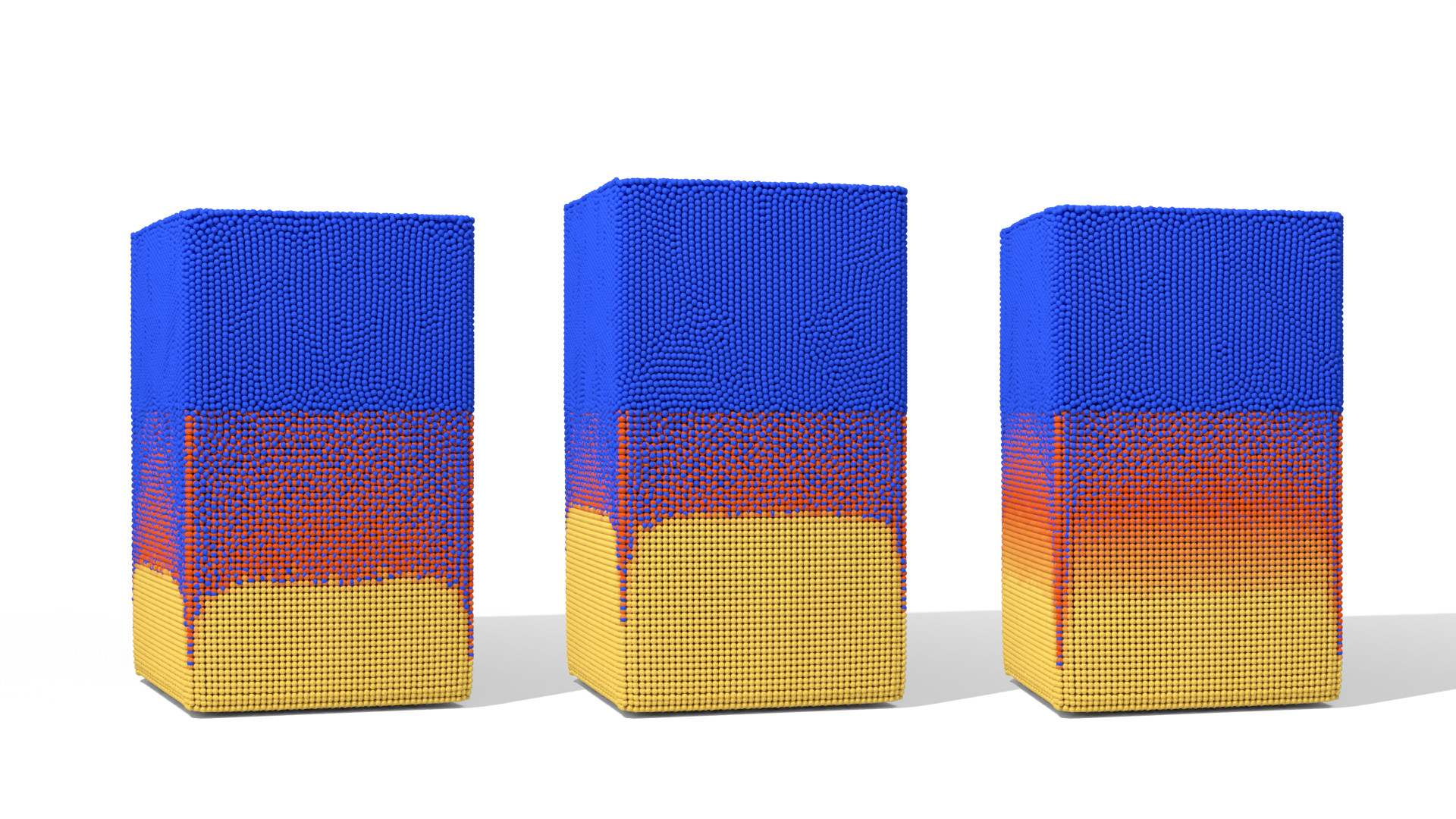}
\caption{Fluid (blue) seeping into porous material, where the particle colors range from no saturation (yellow) to fully saturated (red). Left has a lower viscous drag ($\mu^\text{por} = 10 \si{\pascal \second}$) than the middle ($\mu^\text{por} = 50 \si{\pascal \second}$), resulting in faster seepage. The right has the same viscous drag as the middle, but also adhesion ($C^{\text{cap},0} = 2500 \si{\newton \per \meter}$, $\eta_{\text{cap}} = 0.7$), resulting in faster seepage and a softer transition zone.}
\label{fig:seep}
\end{figure}

To demonstrate the effect of our model parameters, we simulate fluid seeping down into a porous block.
In particular, we are interested in the influence of the porosity $\phi$, the viscous drag coefficient $\mu^\text{por}$ and capillary action controlled by $C^\text{cap,0}$.
First, we analyze the porosity (see \fig{fig:seep_porosity}), which we vary between $\phi \in \{ 0.4, 0.6, 0.8 \}$.
The final state showcases the expected result that the amount of absorbed liquid is linearly proportional to the porosity value.

Next, we investigate the effects of viscous drag and capillary action, while keeping the porosity ($\phi = 0.5$) constant.
The results in \fig{fig:seep} show how we can control the uniform seeping speed using the viscous drag coefficient $\mu^{\text{por}}$, which slows down fluid particles when they interact with solid particles.
The capillary force on the other hand introduces a pull towards unsaturated regions, which increases fluid velocity and leads to a smoother transition between dry and wet regions.
Of these simulations, which each use 110.6k fluid and 110.6k solid particles, the version with adhesion took the longest to compute with an average time of $373.7 \si{\milli \second}$ per time step.
Here, $69 \%$ of the time was spent for the pressure solver and $20 \%$ to compute the strongly coupled non-pressure forces.

\begin{figure}
	\includegraphics[width=\linewidth]{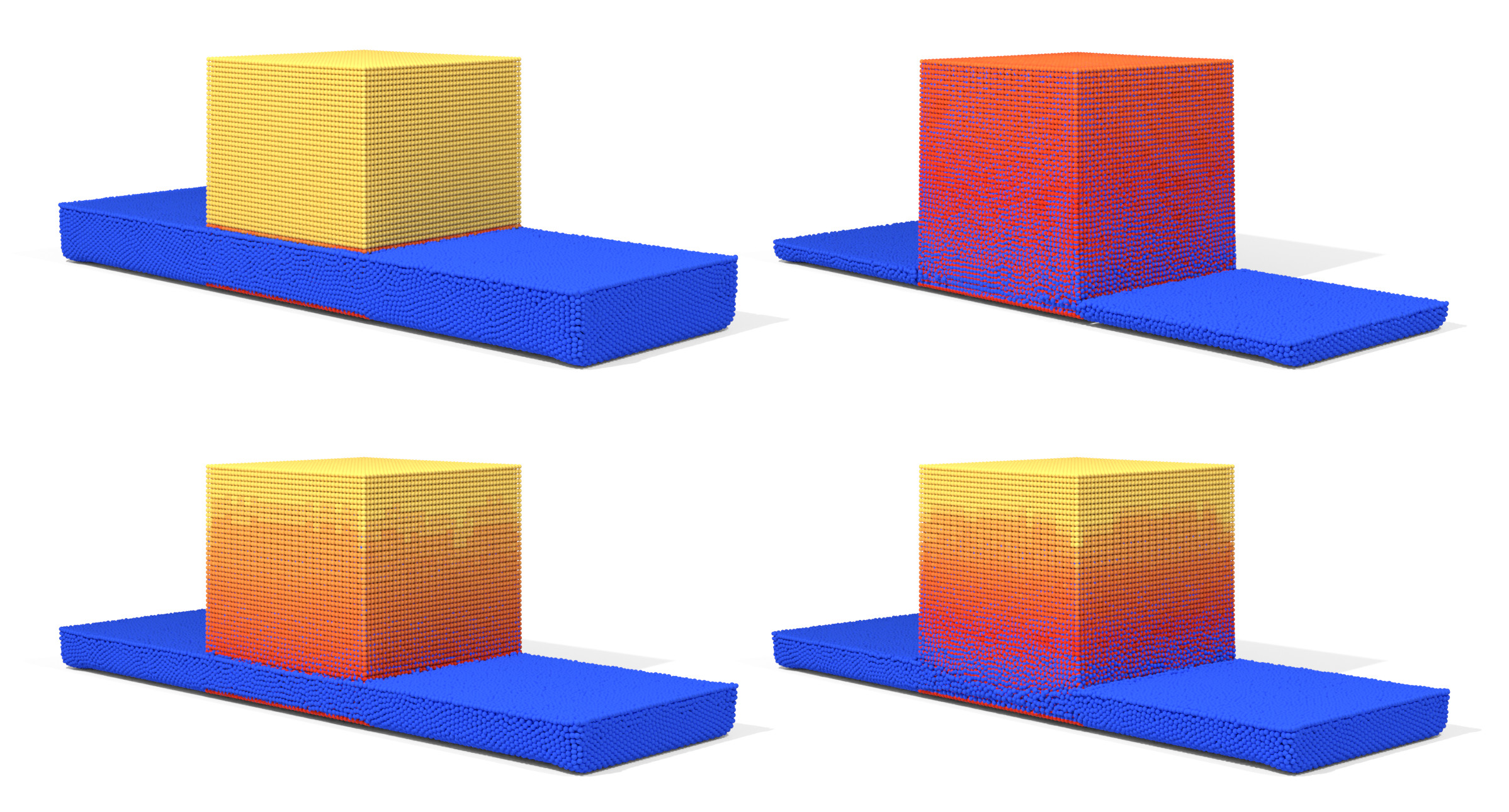}
	\caption{Block with different capillary parameters, after 10\si{\second}. Top left: no capillary action ($C^{\text{cap},0} = 0 \si{\newton \per \meter}$), top right: large adhesion force ($C^{\text{cap},0} = 2500 \si{\newton \per \meter}, \eta_{\text{cap}} = 0.7$). Bottom row: medium adhesion force ($C^{\text{cap},0} = 500 \si{\newton \per \meter}$) with large falloff (left, $\eta_{\text{cap}} = 1$) and reduced falloff (right, $\eta_{\text{cap}} = 0.5$).}
	\label{fig:adhesion}
\end{figure}
To show the effect of the capillary force parameters, we place a cube ($\phi = 0.6$) between two shallow pools of water, shown in \fig{fig:adhesion}.
By varying the capillary force parameter $C^\text{cap,0}$ we can influence both the speed and final height of the absorption.
We first show the simulation without capillary effect ($C^{\text{cap},0} = 0 \si{\newton \per \meter}$), where the fluid permeates only the bottom of the porous medium.
Increasing the capillary force pulls the fluid upwards, such that it can reach the top of the object.
We also show that the falloff $\eta_\text{cap}$, which determines how much the capillary potential diminishes with saturation, controls how much fluid is pulled up into the lower regions of the cube (see also \fig{fig:adhesion_falloff}), while the total height of the wet region only depends on $C^{\text{cap},0}$.

\subsection{Visual Showcases}

\begin{figure}
	\begin{subfigure}{.495\linewidth}
		\includegraphics[width=\linewidth]{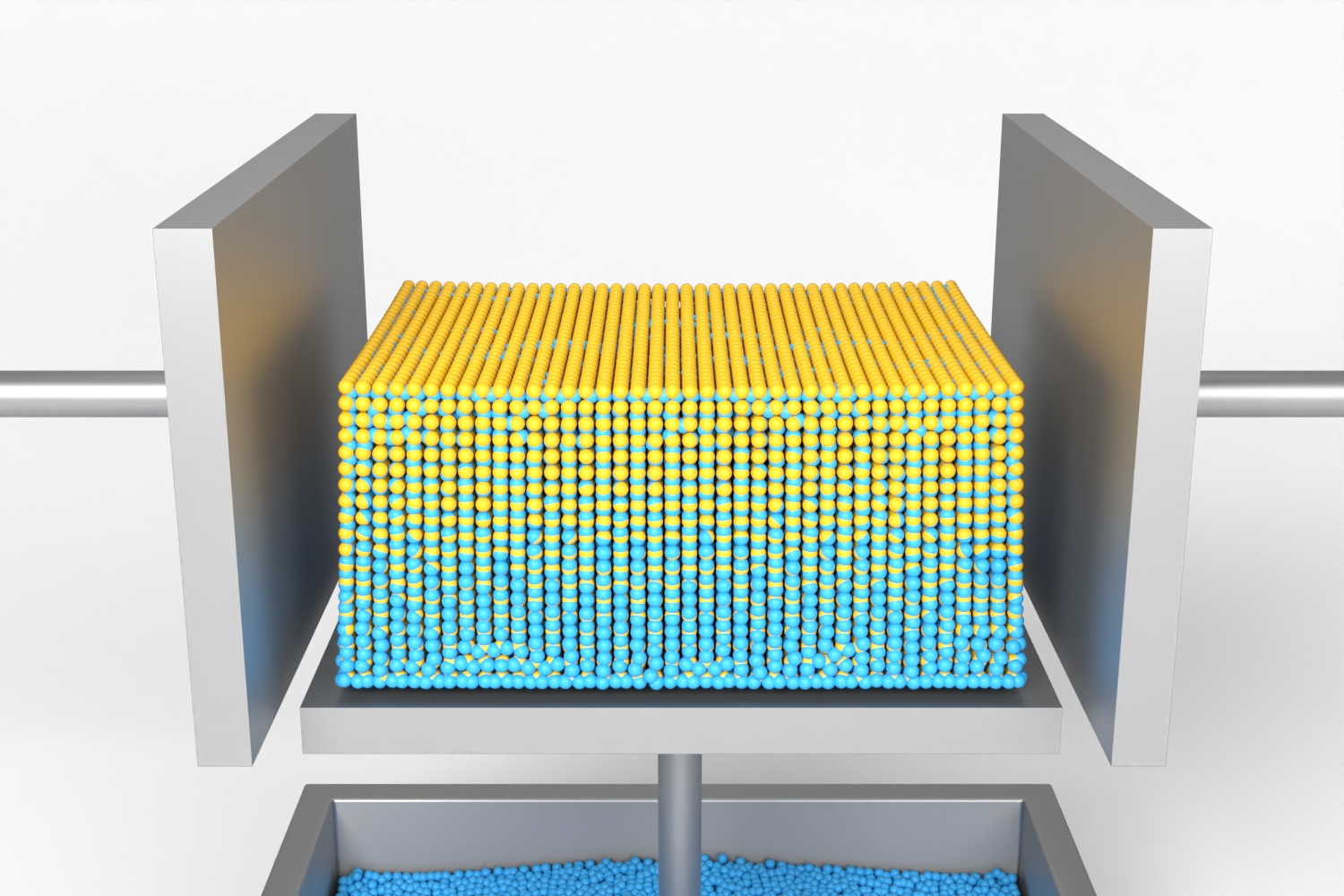}
	\end{subfigure}
	\begin{subfigure}{.495\linewidth}
		\includegraphics[width=\linewidth]{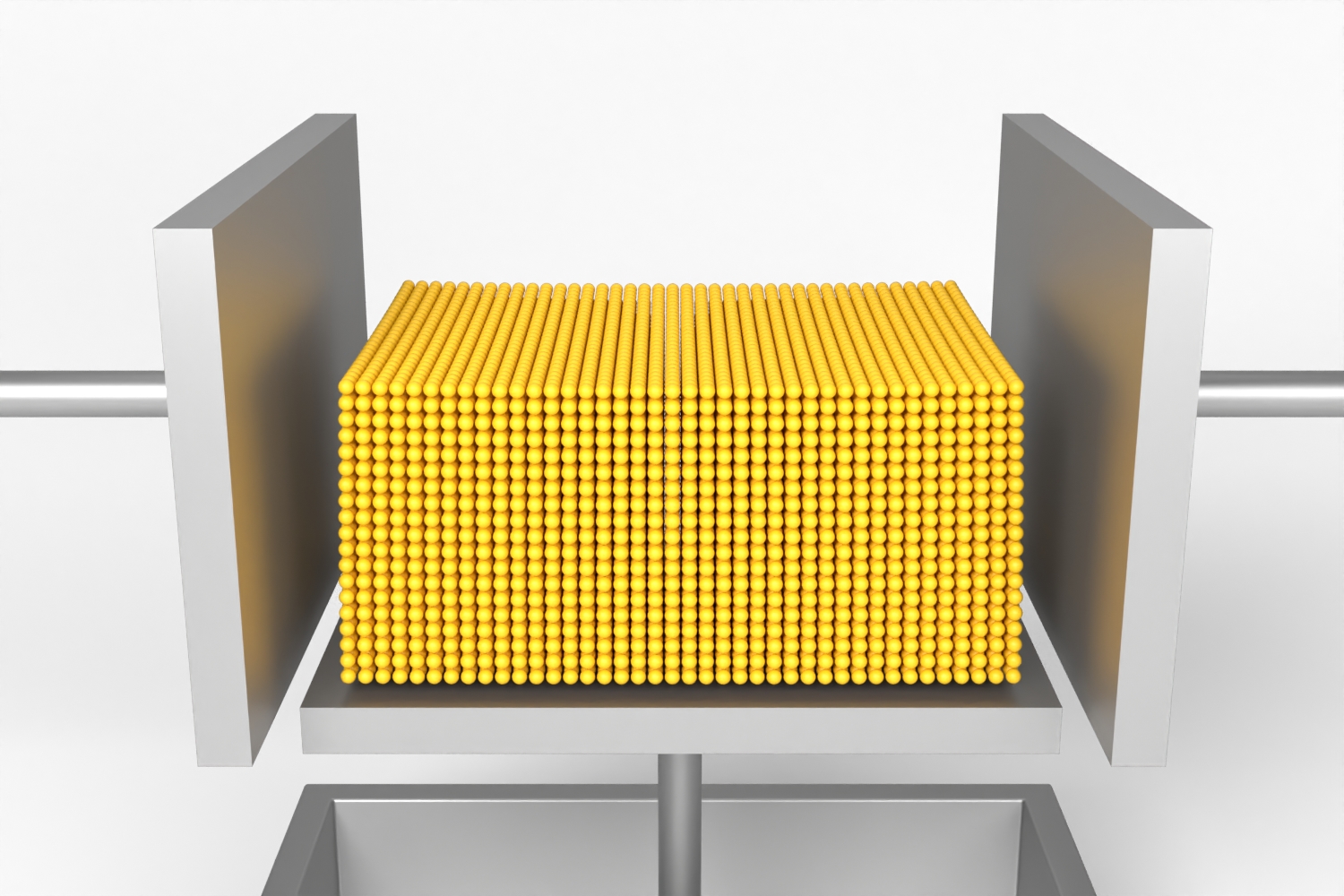}
	\end{subfigure}
	\begin{subfigure}{.495\linewidth}
		\includegraphics[width=\linewidth]{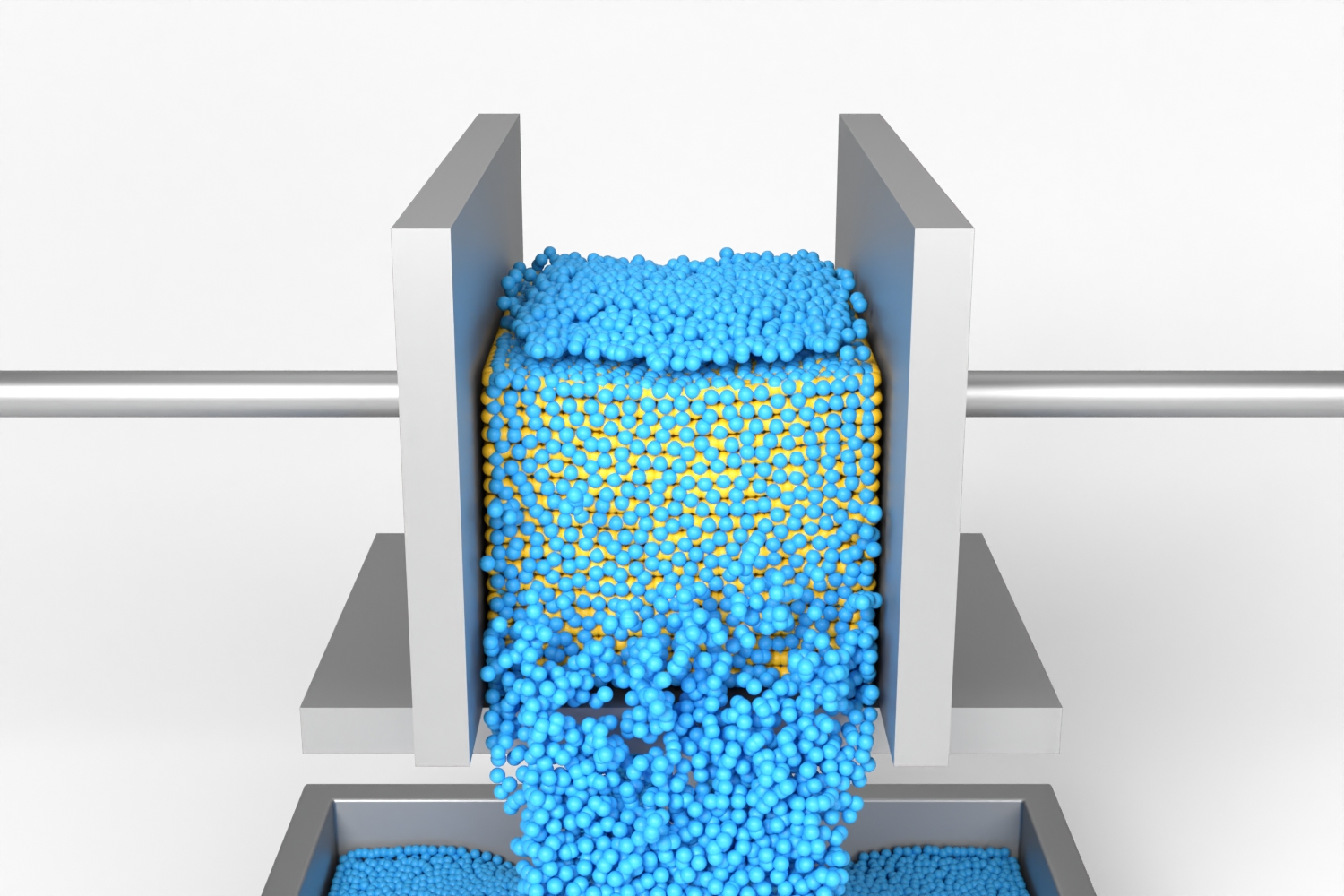}
	\end{subfigure}
	\begin{subfigure}{.495\linewidth}
		\includegraphics[width=\linewidth]{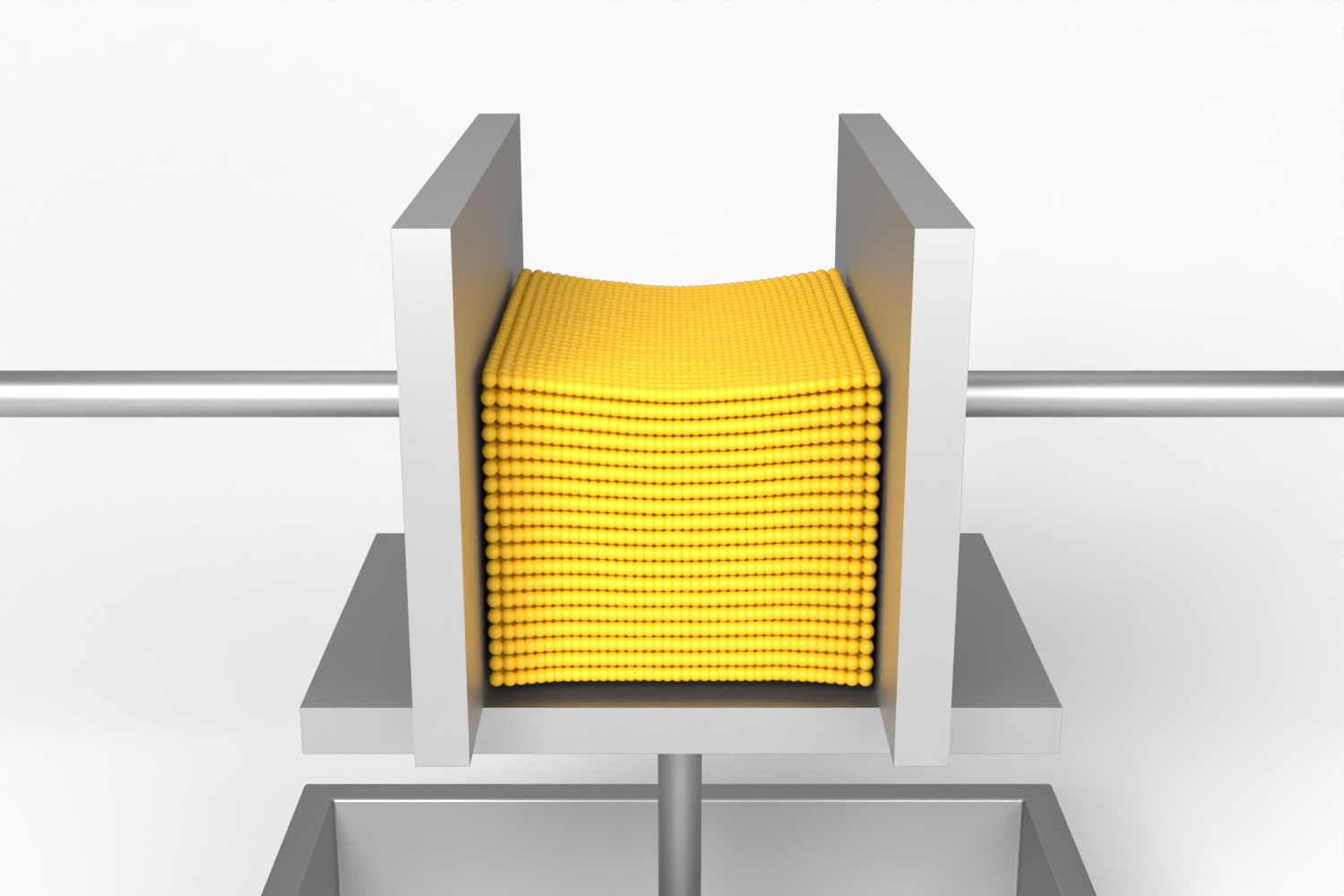}
	\end{subfigure}
	\caption{A saturated sponge being squeezed. Left column shows both solid (yellow) and fluid (blue) particles. The right column only shows the solid particles, which are packed more densely under compression (lower row).} 
	\label{fig:sponge}
\end{figure}

Lastly, we demonstrate that our method can produce complex simulation results with great imaginative liberty.
Shown in \fig{fig:teaser} is the simulation of a compressible sponge,
which is first dropped into a water bath and then subjected to external compression.
We cause the sponge to increase its volume by setting the bloating factor to $\eta_\text{bloat} = 0.01$.
The compression causes solid particles to move closer together, as shown in Fig.~\ref{fig:sponge}, and the corresponding reduction in available pore space causes the absorbed water to be ejected.

\begin{figure*}
	\includegraphics[height=.563\columnwidth,trim={0 0 900 00},clip]{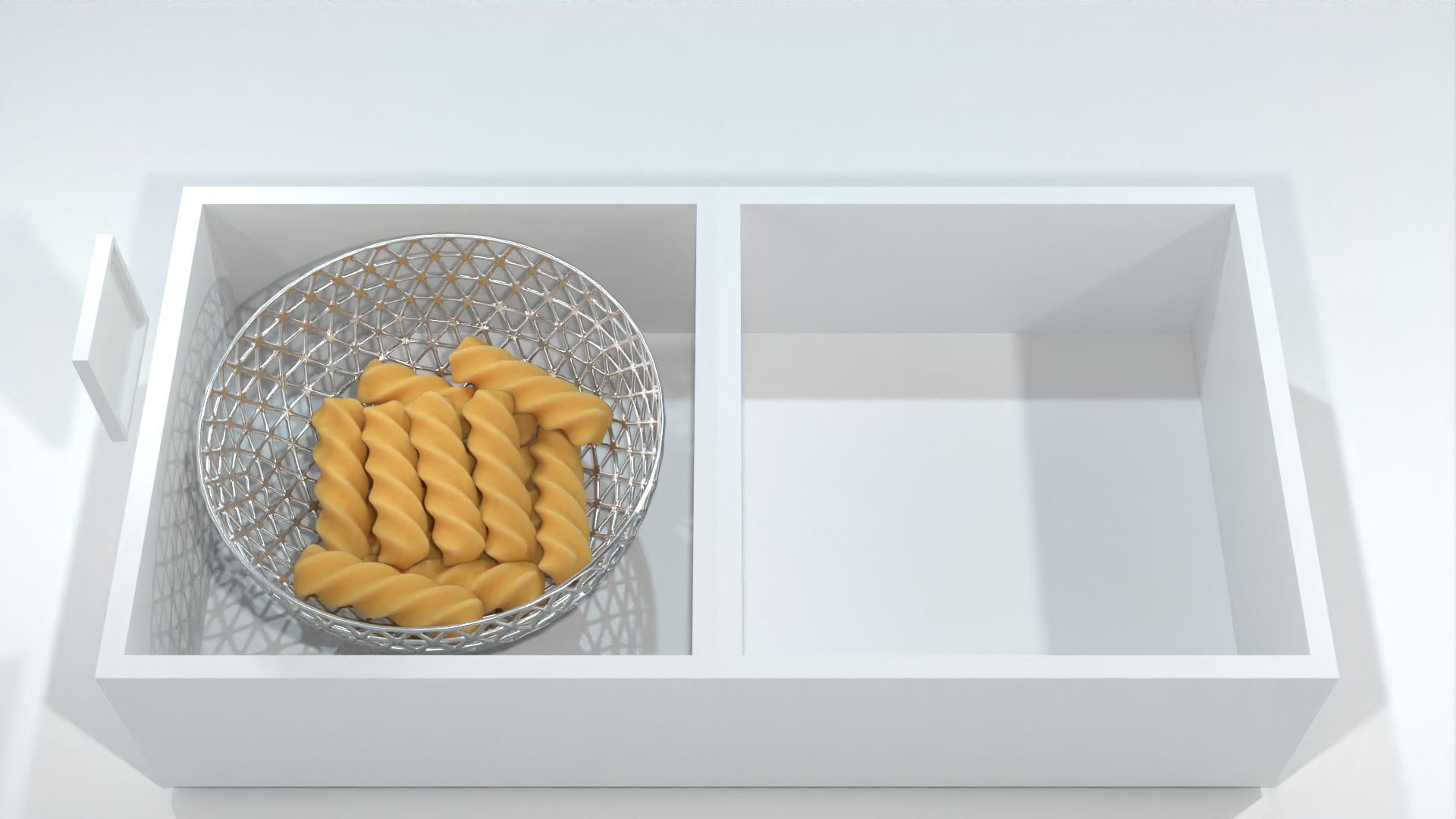}
	\hfill
	\includegraphics[height=.563\columnwidth,trim={0 0 900 00},clip]{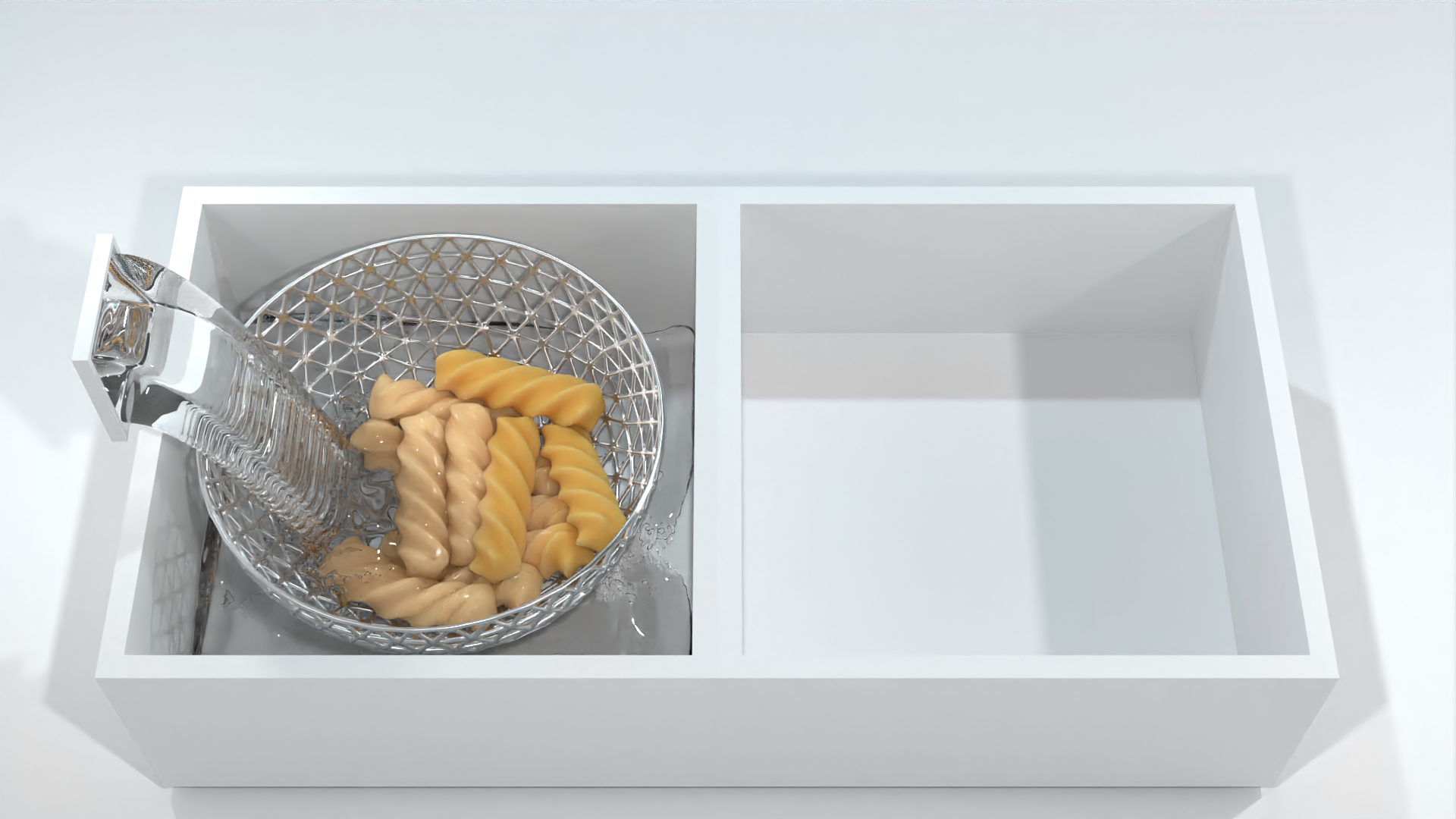}
	\hfill
	\includegraphics[height=.563\columnwidth,trim={0 0 0 00},clip]{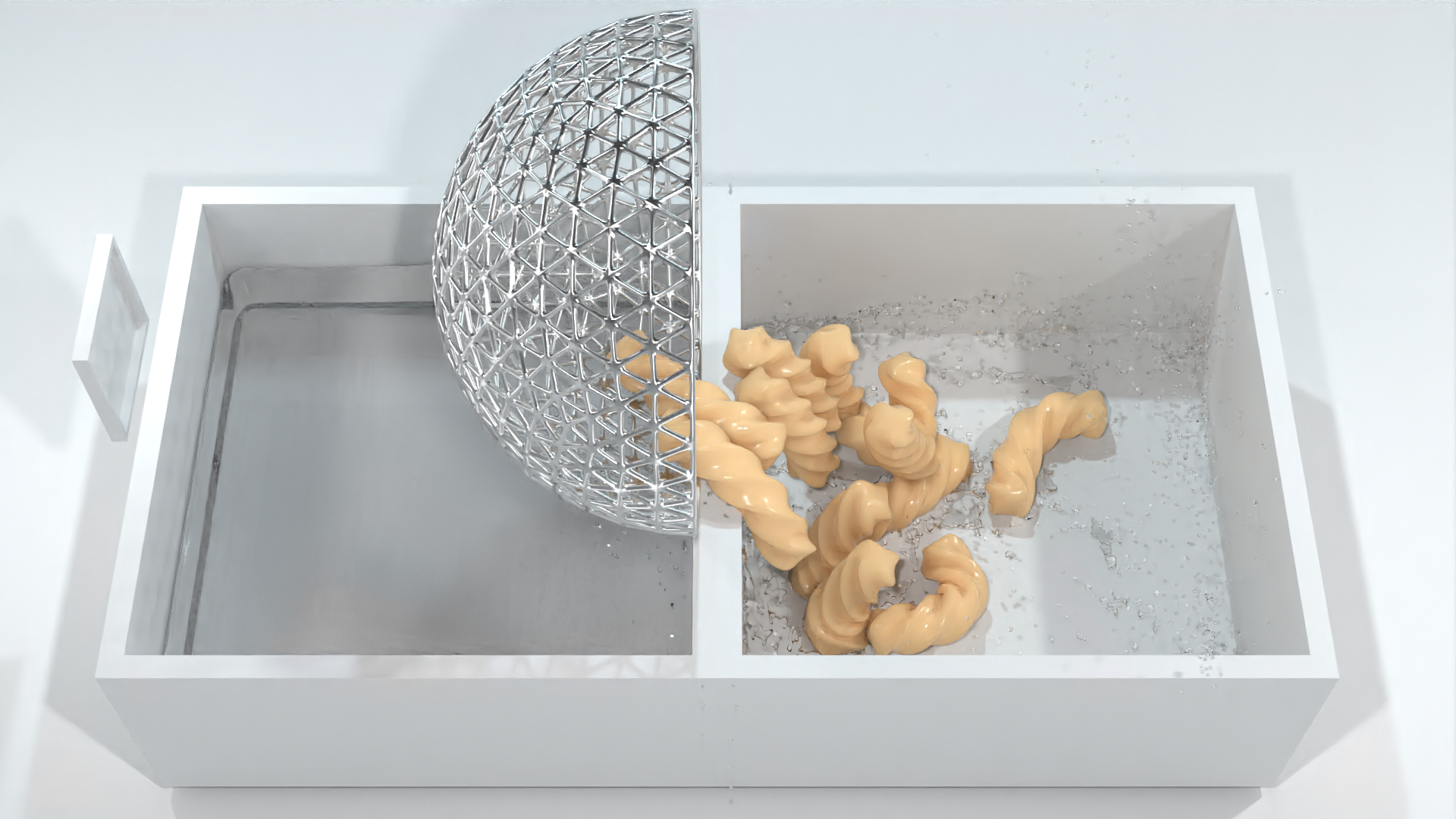}
	\caption{Fusilli-shaped porous objects are submerged in water. The material is initially very stiff, but softens and changes color as it absorbs water.}
	\label{fig:fusilli}
\end{figure*}

\fig{fig:fusilli} shows fusilli-shaped porous objects 
being submerged in water and then spilled into a container.
The objects are initially very stiff, but
setting the shear resistance change factor $\eta_m = -2.0$, while ensuring $\mu(S_s) \geq 0.001 \mu^0$, causes the material to become soft as it absorbs water.

\begin{figure*}
	\includegraphics[width=\linewidth,trim={350 120 310 180},clip]{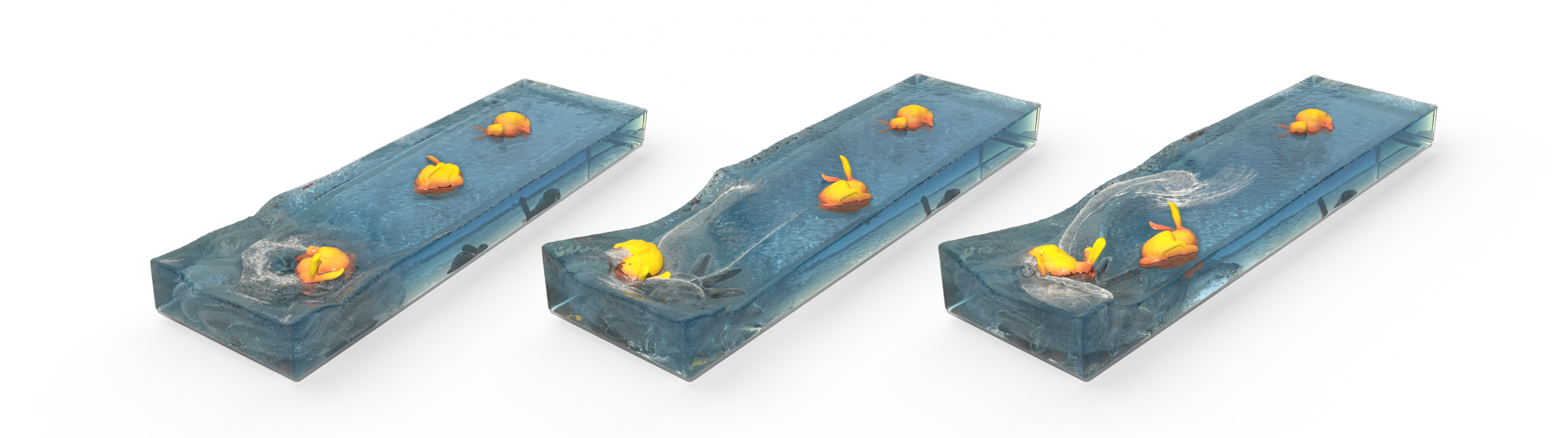}
	\caption{Three porous Stanford Bunnies interact with highly turbulent fluid, accelerated by a spinning propeller submerged under the surface. The bunnies change color from yellow (dry) to orange (wet) based on fluid saturation.}
	\label{fig:vorticity}
\end{figure*}

For simulations that require more turbulent behavior, our solver can also be used in conjunction with SPH fluid vorticity methods.
For the simulation in \fig{fig:vorticity} we use the micropolar model by \citet{Bender19b} to better capture the behavior of fluid accelerated by a rotating propeller. 
Three porous bunnies with a low density are dropped into the fluid bath, where they interact with the surrounding flow field, creating interesting dynamic movements.

\begin{figure}
	\includegraphics[width=\columnwidth,trim={0 300 0 450},clip]{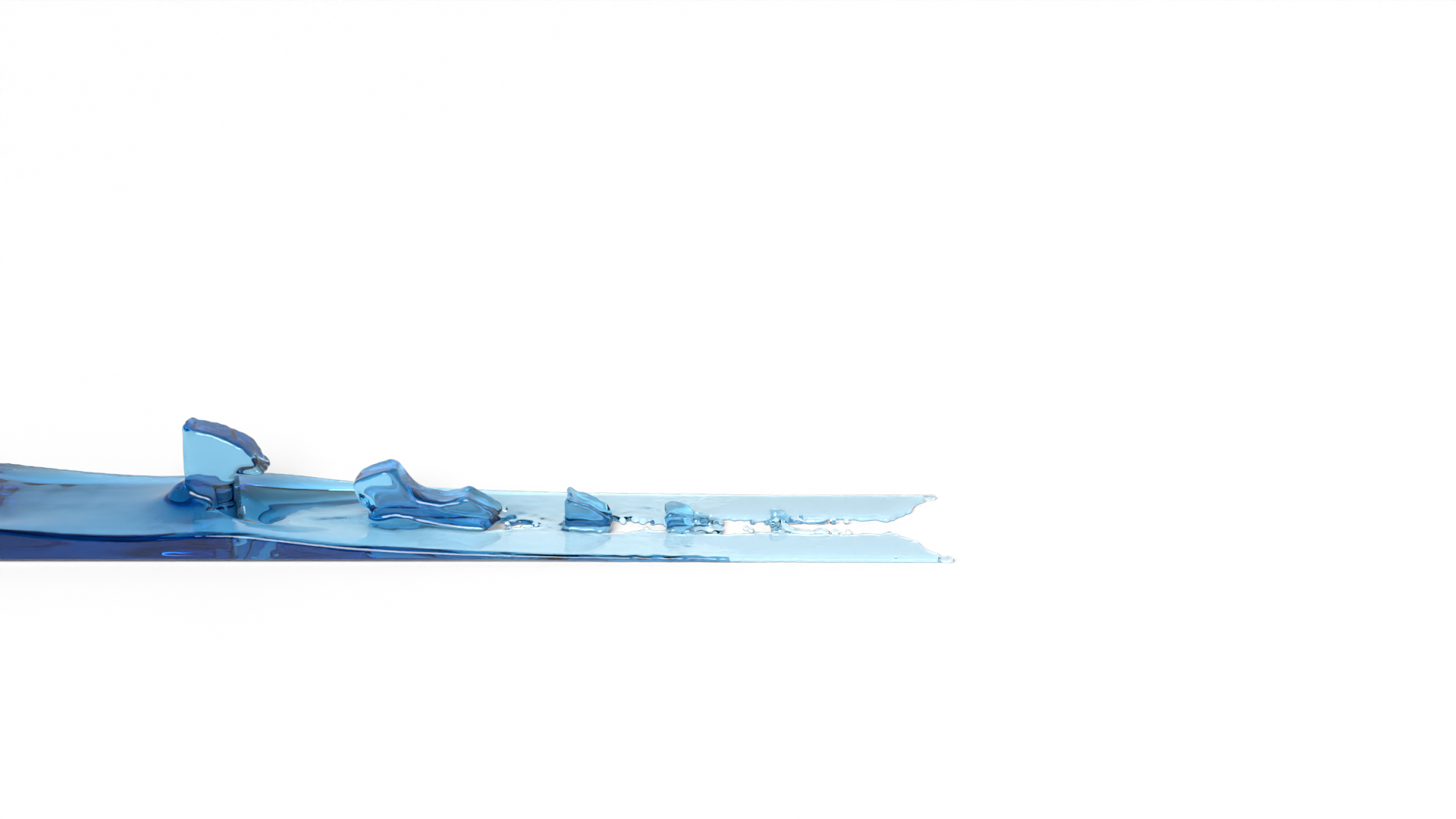}
	\includegraphics[width=\columnwidth,trim={0 300 0 450},clip]{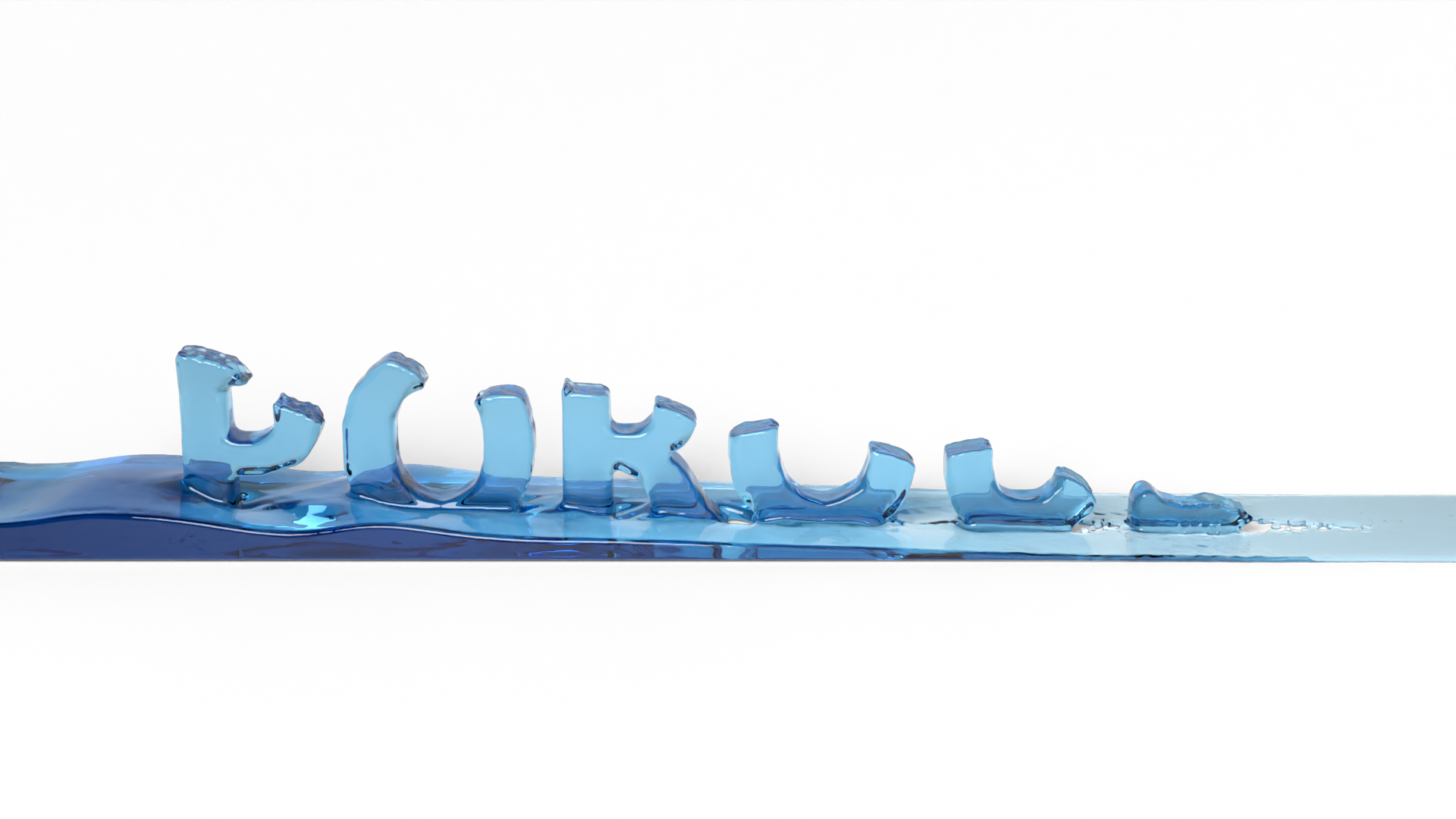}
	\includegraphics[width=\columnwidth,trim={0 300 0 450},clip]{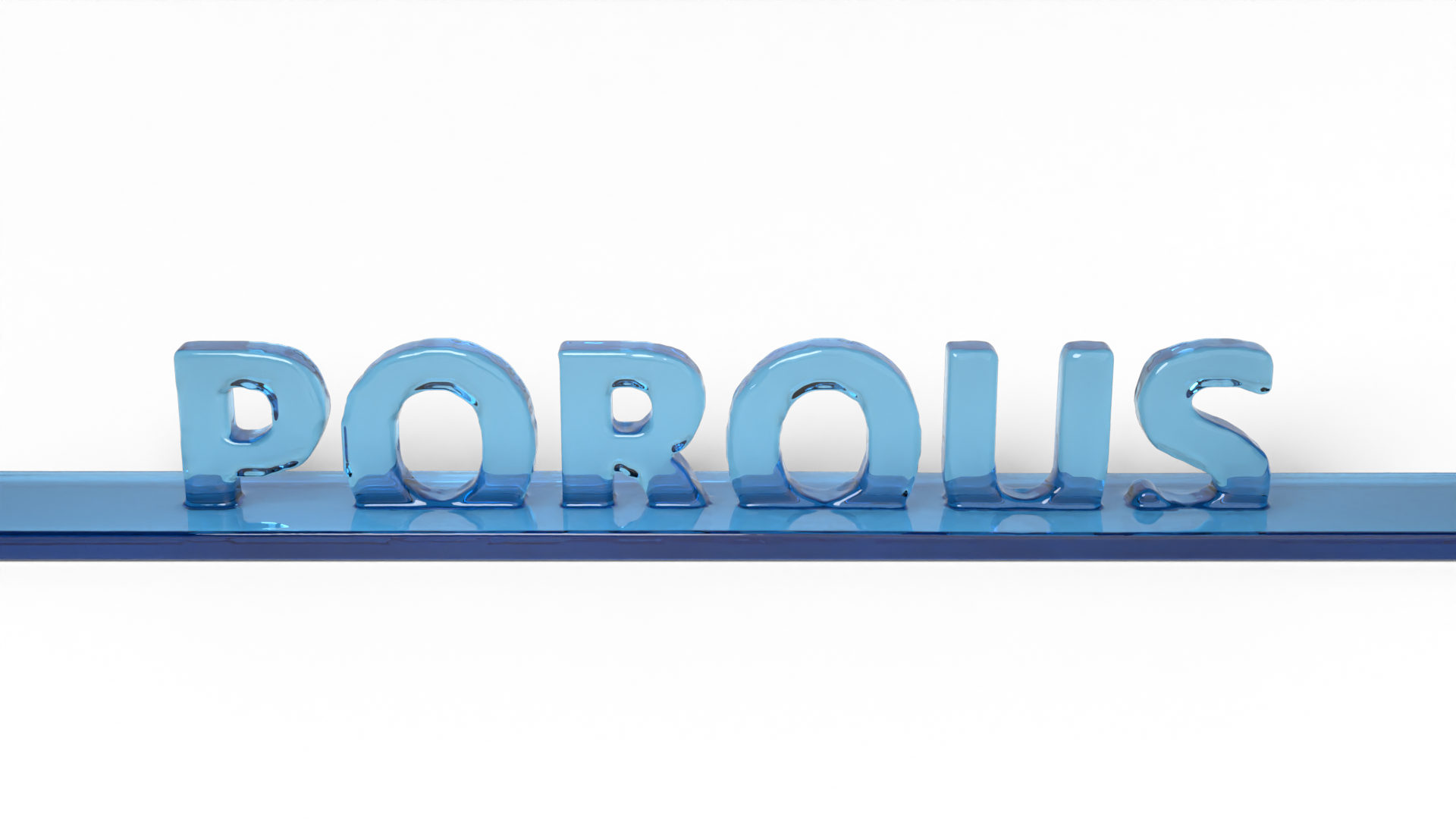}
	\caption{Porous objects in the shape of letters are fixed in space and absorb fluid, which flows in from the left. Here, we only render the fluid surface to show how the fluid is transported upwards by capillary action.}
	\label{fig:porous}
\end{figure}

We also show capillary action with a more artistic scene, using letters made of porous material (see \fig{fig:porous}).
The solid is fixed in space, while the fluid enters from the left and is transported up into the letters through capillary action.

\section{Conclusion}

While previous SPH porous flow methods in computer graphics suffer from stability issues caused by severe runtime particle scaling, our method not only avoids these issues entirely, but further increases stability with the use of implicit formulations and strong coupling of non-pressure forces.
We achieve this by considering porosity within the pressure solver, which allows solid and fluid particles to overlap while still taking incompressibility into account.
We can simulate scenarios that feature the most common porous flow effects, employing viscous drag forces and capillary action due to adhesion.
Overall, our SPH porous flow method not only improves on the current state of the art in terms of stability, but also in physical soundness, due to the use of momentum conserving interaction forces and consistent pressure computations.

Still, our solver has some limitations.
So far, we only incorporated corotated linear elastic solids.
Investigating the use of a different solver for the solid phase is an interesting course of further research, like the use of a non-linear elasticity solver (e.g., \cite{Kee23}) or one for granular materials (e.g., \cite{IWT13}).
We believe our approach of coupling fluid and solid phase has great potential to include even more porous materials.
Additionally, we have not yet investigated non-linear drag models, as we already managed to achieve plausible results with the linear one.
While non-linear drag forces would further increase the physical soundness of our model, deriving an efficient non-linear solver is not trivial and therefore remains as future work.

In our experiments, we observed that the pressure solver is a bottleneck regarding simulation speed.
Using a more efficient pressure solver like divergence-free SPH \citep{Bender15} could increase our performance, but requires further modifications to allow intended velocity divergences occurring in porous flow absorptions and ejections.

\bibliography{bibliography}

%%%%%%%%%%%%%%%%%%%%%%%%%%%%%%%%%%%%%%%%%%%%%%%%%%%%%%%%%%%%%%%%%%%%%%%%%%%%%%%
%%%%%%%%%%%%%%%%%%%%%%%%%%%%%%%%%%%%%%%%%%%%%%%%%%%%%%%%%%%%%%%%%%%%%%%%%%%%%%%
% APPENDIX
%%%%%%%%%%%%%%%%%%%%%%%%%%%%%%%%%%%%%%%%%%%%%%%%%%%%%%%%%%%%%%%%%%%%%%%%%%%%%%%
%%%%%%%%%%%%%%%%%%%%%%%%%%%%%%%%%%%%%%%%%%%%%%%%%%%%%%%%%%%%%%%%%%%%%%%%%%%%%%%
% \newpage
% \appendix
% \onecolumn

% \input{sections/appendix.tex}

%%%%%%%%%%%%%%%%%%%%%%%%%%%%%%%%%%%%%%%%%%%%%%%%%%%%%%%%%%%%%%%%%%%%%%%%%%%%%%%
%%%%%%%%%%%%%%%%%%%%%%%%%%%%%%%%%%%%%%%%%%%%%%%%%%%%%%%%%%%%%%%%%%%%%%%%%%%%%%%

\end{document}